\documentclass[pra,twocolumn,showpacs,preprintnumbers,amsmath,amssymb]{revtex4}

\usepackage{graphicx}
\usepackage{dcolumn}
\usepackage{bm}

\begin{document}


\title{Theoretical studies of high-harmonic generation: Effects
of symmetry, degeneracy and orientation.}

\author{C. B. Madsen}
\author{L. B. Madsen}\affiliation{%
Lundbeck Foundation Theoretical Center for Quantum System Research, Department of Physics and Astronomy, University of Aarhus, 8000 Aarhus C, Denmark.
}%

\date{\today}

\begin{abstract}
Using a quantum mechanical three-step model we present numerical
calculations on the high-harmonic generation from four polyatomic
molecules. Ethylene (C$_2$H$_4$) serves as an example where orbital
symmetry directly affects the harmonic yield. We treat the case of
methane (CH$_4$) to address the high-harmonic generation resulting
from a molecule with degenerate orbitals. To this end we illustrate
how the single orbital contributions show up in the total
high-harmonic signal. This example illustrates the importance of
adding coherently amplitude contributions from the individual
degenerate orbitals. Finally, we study the high-harmonic generation
from propane (C$_3$H$_8$) and butane (C$_4$H$_{10}$). These two
molecules, being extended and far from spherical in structure,
produce harmonics with non-trivial orientational dependencies. In
particular, propane can be oriented so that very high-frequency
harmonics are favorized, and thus the molecule contains prospects
for the generation of UV attosecond pulses.
\end{abstract}


\maketitle
\section{Introduction}\label{Introduction}
In the process of high-harmonic generation (HHG) high-frequency
coherent radiation is emitted from an atom or molecule when exposed
to intense femtosecond laser light. This process has been the
subject of extensive studies during the last couple of decades due
to possible applications, including, e.g., generation of coherent UV
attosecond pulses~\cite{Sansone06,carrera06,cao07}, carrier-envelope
phase retrieval~\cite{haworth07} and tomographic reconstruction of
molecular orbitals~\cite{itatani04,torres1,patchkovskii07}.

Whereas early HHG studies focused on atomic systems, it was not
until the new millennium that experimental work on molecules
gathered pace. Studies on HHG from molecules is motivated by the
expectation that the more degrees of freedom and non-spherical
symmetry as compared to atoms may lead to richer physics and a
higher degree of control. Hitherto the molecules considered have
mostly been diatomic or linear systems and it is only just recently
that experimental results on HHG from more complicated molecules
have become available~\cite{altucci1}. Here we apply a quantum
mechanical three-step model to investigate theoretically HHG from
polyatomic molecules of current interest to
experimentalists~\cite{altucci1,torres1}. In our theory, we use a
detailed description of the molecular orbitals, obtained from
Hartree-Fock calculations. We illustrate the effect of orbital
symmetry on HHG by studying the orientation dependence of the
harmonic signal from ethylene (C$_2$H$_4$). The methane molecule
(CH$_4$) is used to demonstrate how degenerate HOMOs influence the
harmonic spectrum. In particular, we show the importance of
including coherently the contribution from every single HOMO when
calculating the harmonic yield from an oriented molecule. We
demonstrate how the extended and non-spherical molecules propane
(C$_3$H$_8$) and butane (C$_4$H$_{10}$) give rise to harmonic
spectra with a rather complex orientational dependence. In the case
of propane harmonics near the cutoff can be selected by orienting
the molecule, and since we expect such harmonics to be
synchronically emitted~\cite{lu2006} orientation of propane is
identified as a tool for generating attosecond pulses. It is the
first time, to our knowledge, an investigation of the detailed
orientational dependence of the single molecule high-harmonic signal
from such complex systems is presented.

The paper is organized as follows. In Sec.~\ref{Theory} we derive
formulas for the single-particle respons from a system with
degenerate HOMOs, and we review the quantum mechanical three-step
model used to calculate the harmonic yield~\cite{madsen2}. In
Sec.~\ref{CalcDet} we briefly discuss some numerical details for our
calculations. Section~\ref{Results} contains numerical results
describing the harmonic yield from each of the aforementioned
molecules. Especially, we treat several effects related to the
orientation of such systems. Finally, we give a summary and the
conclusions in Sec.~\ref{Conclusion}. [We use atomic units
$(e=\hbar=m_e=a_0=1)$ throughout.]

\section{Theory}\label{Theory}
\subsection{Harmonic yield from a statistical mixture of
molecules}\label{TheoryA} The complex amplitude for the emission of
harmonics with frequency $\omega_\text{HHG}$ polarized along the
linear polarization vector $\bm{e}$ is obtained from the Fourier
transform of the dipole acceleration
\begin{equation}\label{A}
A_{\bm{e}}(\omega_\text{HHG})  = \bm{e} \cdot \int dt\, e^{-i
\omega_\text{HHG} t} \frac{d^2}{dt^2} \langle \hat{\bm{d}} \rangle
(t),
\end{equation}
where $ \langle \hat{\bm{d}} \rangle (t)$ is the expectation value
of the dipole operator $\hat{\bm d}$ of the molecule. The
corresponding power density is given by \cite{MilonniD,BurnettHHG}:
\begin{equation}\label{spectrum}
S_{\bm{e}}(\omega_\text{HHG}) \propto \vert
A_{\bm{e}} (\omega_\text{HHG}) \vert^2.
\end{equation}
The quantum state will be a mixed state, since several unobserved
variables appear in a measurement of the harmonic signal. First,
typically several electrons from a given molecule will contribute
significantly to the HHG because of degeneracy of the highest
occupied molecular orbital (HOMO). An experiment recording the total
harmonic yield from an ensemble of ground state molecules cannot
distinguish contributions from the individual degenerate HOMOs.
Second, since perfect orientation of molecules is unrealistic, the
measured harmonic yield arises from an ensemble of molecules with
different orientations. In this section we treat these unresolved
degrees of freedom, using and elaborating the results of recent
work~\cite{madsen1}. Initially, the molecule is in a stationary
thermal state at temperature $T$, and the system is completely
characterized by the density matrix $\hat{\rho}_0 =
\exp{(-\hat{H}/k_B T)}/Z$, with partition function $Z =
\textnormal{Tr}[\exp(-\hat{H}/ k_B T)]$, $\hat{H}$ the field-free
molecular Hamiltonian and $k_B$ Boltzmann's constant. We want to
resolve the molecular initial state on energy eigenstates. The
molecules we consider are, however, so complex that several
simplifying assumptions have to be made in order to obtain a
practical theoretical formulation. According to the Born-Oppenheimer
approximation, we can separate the electron and nuclear motion, and
we will assume only the electronic ground state be populated. Next
we adapt the single-active-electron (SAE) approximation, and
introduce an index $\lambda$ in order to be able to discriminate
between the degenerate HOMOs of this active electron. The nuclear
motion consists of rotation and vibration. We shall assume that only
the vibrational groundstate is occupied, whereas the rotation needs
to be treated in more detail to take account of an oriented
molecule. In doing so, we neglect the rovibrational interaction such
that rotation and vibration can be treated separately. The rotation
is characterized by the asymmetric top quantum numbers $J,\tau$ and
$M$~\cite{zare1}. We may, accordingly, specify the energy
eigenstates by their electronic and rotational degrees of freedom,
$\vert\lambda\rangle\otimes\vert J\tau M\rangle$. When the system
interacts with a laser, the energy eigenstates will evolve according
to a unitary operator that describes any number of orienting pump
pulses followed by a short, intense laser pulse that drives the HHG
[spontaneous decay processes can be neglected on the timescales we
consider]. The orientation pulses are not strong enough to affect
the electronic motion appreciably~\cite{stapelfeldt2003}, while the
nuclear dynamic can be considered as frozen during the short
high-harmonic generating pulse~\cite{madsen1}. Thus, if the delay
between the final orienting pulse and the driving pulse is denoted
by $t_d$, we can split the time evolution operator according to
$U(t)\simeq U_{\text{HHG}}(t)\otimes U_{\text{orient}}(t_d)$ for a
description of the evolution when the HHG is produced, where
$U_\text{HHG}(t)$ propagates the electronic part of the molecular
state during the pulse of the driving laser and
$U_\text{orient}(t_d)$ accounts for the evolution of the rotational
state of the molecule. The energy eigenstates then evolve as follows
\begin{eqnarray}\label{eq:timedevolution}
\vert\Psi_{J\tau
M}^\lambda(t)\rangle&\simeq&\left(U_{\text{HHG}}(t)\vert\lambda\rangle\right)
\otimes\left(U_{\text{orient}}(t_d)\vert J\tau
M\rangle\right)\nonumber \\
&=&\vert\psi_\lambda(t)\Phi_{J\tau M}(t_d)\rangle.
\end{eqnarray}
We return to the time evolution of the system in
Sec.~\ref{timedevolution}. Our current goal is to calculate the
expectation value of the dipole operator as this enters
Eqs.~\eqref{A} and~\eqref{spectrum}. This evaluation is most
conveniently done by expanding the energy eigenstates in the
position basis $\vert\bm{r},\phi,\theta,\chi\rangle$ in which the
dipole operator is diagonal
\begin{eqnarray}\label{eq:dipolemean}
\langle\hat{\bm{d}}(t)\rangle=\text{Tr}\left[\hat{\rho}(t)\hat{\bm{d}}\right]
=\text{Tr}\left[U(t)\hat{\rho}_0U^\dagger(t)\hat{\bm{d}}\right]\simeq\nonumber
\\
\int_0^{2\pi} d\phi \int_0^\pi d\theta \sin\theta \int_0^{2\pi}
d\chi
G_{t_d}(\phi,\theta,\chi)\sum_\lambda\langle\bm{\hat{d}}_\lambda\rangle(\phi,\theta,\chi,t)
\end{eqnarray}
with
\begin{eqnarray}
\langle\bm{\hat{d}}_\lambda\rangle(\phi,\theta,\chi,t)&=&\int
d\bm{r}
\vert \psi_\lambda(\bm{r},\phi,\theta,\chi,t)\vert^2\bm{\hat{d}}, \\
G_{t_d}(\phi,\theta,\chi) &=& \sum_{J,\tau, M} P_{J\tau}\vert
\Phi_{J\tau M}(\phi,\theta,\chi;t_d)\vert^2,\label{eq:Gtd}
\end{eqnarray}
and $P_{J\tau}=\exp(-E_{J\tau}/k_BT)$ the Boltzmann weight of the
asymmetric top wavefunction $\langle\phi,\theta,\chi \vert
U_{\text{orient}}(t_d)\vert J\tau M\rangle=\Phi_{J\tau
M}(\phi,\theta,\chi;t_d)$ of energy $E_{J\tau}$.

Consequently, using Eqs.~\eqref{A} and~\eqref{spectrum}, the
harmonic signal is given by
\begin{eqnarray}\label{eq:signal}
S_{\bm{e}}(\omega_\text{HHG})\propto\Bigg|\sum_\lambda\int_0^{2\pi}
d\phi \int_0^\pi d\theta \sin\theta\int_0^{2\pi} d\chi
\nonumber \\
\times
G_{t_d}(\phi,\theta,\chi)A_{\bm{e}}^\lambda(\omega_\text{HHG},\phi,\theta,\chi)\Bigg|^2
\end{eqnarray}
with
\begin{equation}\label{eq:cplx_amp}
A_{\bm{e}}^\lambda(\omega_\text{HHG},\phi,\theta,\chi)=\bm{e}
\cdot\int dt\, e^{-i \omega_\text{HHG} t} \frac{d^2}{dt^2}
\langle\bm{\hat{d}}_\lambda\rangle(\phi,\theta,\chi).
\end{equation}
In general, the degenerate HOMOs will interfere due to the coherent
sum over $\lambda$ in Eq.~\eqref{eq:signal}. Nevertheless, there are
special cases where the degeneracy enters simply as a factor
multiplying the signal from a single HOMO. For example we mention
that if the degeneracy is due to spin multiplicity, $N_\text{S}$,
the sum over different HOMOs yields a factor $N_\text{S}^2$ in the
signal. Another instance occurs when the HOMOs differ simply by a
rotation. Then, in the case of a randomly oriented ensemble of
molecules, i.e., $G_{t_d}=1/(8\pi^2)$, it is obvious that HOMOs must
each give rise to the same complex number when the single HOMO
amplitude, $A_{\bm{e}}^\lambda(\omega_\text{HHG},\phi,\theta,\chi)$,
is averaged over all orientations. It then follows from
Eq.~\eqref{eq:signal} that the degeneracy will again enter as a
factor multiplying the signal from a single HOMO.

\subsection{Model}\label{timedevolution}
As described in the previous section the system starts out in a
statistically mixed state composed of the energy eigenstates. Each
$\vert\lambda\rangle\otimes\vert J\tau M\rangle$ state evolves
according to Eq.~\eqref{eq:timedevolution}. In this section we focus
on the time evolution of the system.

First, propagating the asymmetric top energy eigenstates, $\vert
J\tau M\rangle$, to obtain the $\vert \Phi_{J\tau M}(t_d)\rangle'$s
at the time $t_d$ of the high-harmonic probing pulse is a
numerically demanding task. The issue has already been addressed in
several studies~\cite{stapelfeldt2003,rouzee2006,underwood2005}, and
in the present work, we will assume either $G_{t_d}$ to be simply
uniform (no preferred orientation) or use an idealized orientational
distribution to be specified in Sec.~\ref{Results}.

Next, we focus on the electronic part of the time evolution. In the
field-free initial state the HOMO wave function is conveniently
expressed in a spherical expansion in the body fixed (BF) frame
\begin{equation}\label{eq:HOMO}
\psi_\lambda^{\text{BF}}(\bm{r})=\sum_{l,m} F_{l,m}^\lambda(r)
  Y_l^m(\hat{\bm{r}}).
\end{equation}
Asymptotically this expression must follow the Coulomb form
\begin{equation} \label{atiwave}
  \psi_\lambda^{\text{BF}}(\bm{r})\sim \sum_{l,m} C_{l,m}^\lambda r^{\mathcal{Z} / \kappa -1}
  \exp(-\kappa r) Y_l^m(\hat{\bm{r}})
\end{equation}
with $\kappa = \sqrt{2I_P}$, $I_P$ the ionization potential,
$\mathcal{Z}$ the net charge of the molecule when the HOMO electron
is removed and where the $C_{l,m}^\lambda$'s are fitting
coefficients. More detail on how to obtain wavefunctions in
Eqs.~\eqref{eq:HOMO} and~\eqref{atiwave} is given in
Sec.~\ref{CalcDet}. We wish to carry out calculations in a
laboratory fixed (LF) system defined by the laser polarization, and
for a molecule of arbitrary orientation LF and BF coordinate axes do
not in general coincide. Hence, we rotate the BF wave function to
obtain the LF wave function by application of the rotation operator
\begin{equation} \label{homoex}
  \psi_\lambda^{\text{LF}}(\bm{r},\phi,\theta,\chi)=\hat{D}(\phi,\theta,\chi)\psi_\lambda^{\text{BF}}(\bm{r}),
\end{equation}
where the rotation is given by the Euler angles $\phi$, $\theta$ and
$\chi$. Following the conventions of~\cite{zare1} $\theta$ is the
angle between the BF $z$-axis and the LF $Z$-axis, $\phi$ denotes a
rotation around the $Z$-axis, and finally $\chi$ denotes a rotation
around the $z$-axis. Note that the effect of $\hat{D}$ is readily
evaluated in the spherical harmonic basis used in
Eqs.~\eqref{eq:HOMO} and~\eqref{atiwave}~\cite{zare1}.

We consider the case where the driving laser pulse contains several
cycles such that a Floquet approach is suitable. Hence using the
Coulomb gauge and the dipole approximation, a laser field with
frequency $\omega$ and period $T=2\pi/\omega$ is described by the
vector potential $\bm{A}(t)=\bm{A}_0 \cos(\omega t)$. According to
the quantum mechanical three-step model described
in~\cite{kuchievandostrovsky} for atoms and~\cite{madsen2} for
molecules the electronic time evolution, given by
$U_{\text{HHG}}(t)$ [see Eq.~\eqref{eq:timedevolution}], consists of
a HOMO electron being transfered to the continuum via above
threshold ionization (ATI), i.e., by absorbing a number of photons
from the driving laser. The electron then propagates in the
laser-dressed continuum and is eventually, due to the periodicity of
the laser field, driven back to a recombination with the molecule,
where it returns to the HOMO. Within this model the complex
amplitude for the emission of harmonics polarized along the unit
vector $\bm{e}$ with frequency $\omega_\text{HHG}^N=N\omega$ ($N$
integer) is~\cite{madsen2}
\begin{widetext}
\begin{eqnarray} \label{harmonic}
  A_{\bm{e}}^\lambda(\omega_\text{HHG}^N,\phi,\theta,\chi)\propto
  \sum_{l_2,l_1}\sum_{m_2',m_1'}\sum_{m_2,m_1}D_{m_2',m_2}^{l_2*}
  (\phi,\theta,\chi)D_{m_1',m_1}^{l_1}
  (\phi,\theta,\chi)C_{l_1,m_1}^\lambda
  \sum_{k}\sum_{C(k)}B
  _{l_2,m_2',m_2}^{\lambda,N,k,\bm{e}}(C(k))A_{l_1,m_1'}^{k}
  (C(k)).
\end{eqnarray}
Here $D_{m_i',m_i}^{l}(\phi,\theta,\chi)$ with $i=1,2$ is the Wigner
rotation function~\cite{zare1}, while
\begin{eqnarray}\label{Aamp}
  C_{l_1,m_1}^\lambda A_{l_1,m_1'}^{k}(C(k)) = -C_{l_1,m_1}^\lambda\frac{1}{T}\Gamma
  \left(1+\frac{\mathcal{Z} / \kappa}{2}\right) 2^{\frac{\mathcal{Z} / \kappa}{2}}
  \kappa^{\mathcal{Z} / \kappa}(\pm 1)^{l_1}\frac{\exp [iS(t_{C(k)}')]}
  {\sqrt{[-iS''(t_{C(k)}')]^{1+\mathcal{Z} / \kappa}}}
  \left. Y_{l_1}^{m_1'}\left(\hat{\bm{q}}'\right)\right
  \arrowvert_{\bm{q}'=\bm{K}_k+\bm{A}(t_{C(k)}')}
\end{eqnarray}
and
\begin{eqnarray}\label{Bamp}
  B_{l_2,m_2',m_2}^{\lambda,N,k,\bm{e}}(C(k)) = i \frac{(2\pi)^2}
  { T} \int_0^T dt \frac{\exp[i(N\omega t-S(t))]}{L_0(t,t_{C(k)}')}
  (\bm{e}\cdot \bm{\nabla}_
  {\bm{q}} ) \left. \left[ \tilde{F}_{l_2,m_2}^\lambda(q)Y_{l_2}^{m_2'}
  \left(\hat{\bm{q}}\right) \right]^*\right\arrowvert_{
  \bm{q}=\bm{K}_k+\bm{A}(t)},
\end{eqnarray}
\end{widetext}
along with their Wigner rotation functions, are interpreted as ATI
and propagation-recombination amplitudes, respectively, of a HOMO
electron having absorbed $k$ photons during the ATI-step. In
Eqs.~\eqref{Aamp} and~\eqref{Bamp} $\bm{q}$ and $\bm{q}'$ are
electron momenta and
\begin{equation}\label{action}
  S(t) = k\omega t+\bm{K}_k\cdot \frac{\bm{A}_0}{\omega}
  \sin(\omega t) + \frac{U_p}{2\omega}\sin(2\omega t)
\end{equation}
is the quasi-classical action. The index $C(k)$ in
Eqs.~\eqref{harmonic}-\eqref{Bamp} denotes the saddle-points. For
each $k$ the saddle-points $t'_{C(k)}$ are defined by the condition
$S'(t'_{C(k)})=0$, and we use the ones with $0 \leq
\text{Re}(t_{C(k)}')< T$ along with $\text{Im}(t_{C(k)}')>0$. The
factor $(\pm 1)^{l_1}$ in Eq.~\eqref{Aamp} corresponds to the limits
$\pm i\kappa$ of the size $q'$ of the electron momentum at the
saddle-points. The factor
$1/L_0(t,t'_{C(k)})=\sigma\alpha_0(\sin\omega t'_{C(k)}-\sin\omega
t)$ in Eq.~\eqref{Bamp}, with $\sigma=\pm 1$ to assure
$\text{Re}(L_0)>0$, describes the decrease of the amplitude of the
electron wave as it propagates in the field-dressed continuum. Also,
$\bm{K}_k$ is the part of the continuum electron momentum arising
from absorption of $k$ laser photons during ATI, thus
$K_k=\sqrt{2(k\omega-I_p-U_p)}$ with $U_p=A_0^2/4$ the ponderomotive
potential and $\bm{e}_{\bm{K}_k}=\sigma\bm{e}_{\bm{A}_0}$. Finally,
in Eq.~\eqref{Bamp} the function $\tilde{F}_{l_2,m_2}^\lambda(q)$ is
the radial part of the momentum space HOMO wave function, obtained
by taking the Fourier transform of Eq.~\eqref{homoex}
(see~\cite{madsen2} for further details).

\section{Calculational details}\label{CalcDet}
We have determined the Hartee-Fock wave functions for ethylene
(C$_2$H$_4$), methane (CH$_4$), propane (C$_3$H$_8$) and butane
(C$_4$H$_{10}$) in a spherical basis along with the asymptotic
coefficients using the technique described in~\cite{kjeldsen1}. In
Table~\ref{tab:table1} we list the ionization potentials $I_p$ and
the asymptotic $C_{l,m}^\lambda$-coefficients entering
Eq.~\eqref{atiwave}. Note that the results in the case of ethylene
differ from those in~\cite{kjeldsen1} because our choice of the BF
axes in the present work follows the convention of~\cite{rouzee2006}
with the $xz$-plane coinciding with the molecular plane, whereas
in~\cite{kjeldsen1} the molecular plane was chosen to coincide with
the $yz$-plane.
\begin{table*}
\caption{\label{tab:table1}The molecular properties of the alkalenes
used in this work for evaluation of HHG. $I_p$ is the experimental
adiabatic ionization potential~\cite{NIST}. The remaining numbers in
the table give the values of the asymptotic coefficients,
$C_{l,m}^\lambda$, entering Eq.~\eqref{atiwave} based on GAMESS
calculation~\cite{GAMESS} using a triple zeta valence basis set with
diffuse sp shells. The three degenerate HOMOs in CH$_4$ are denoted
by HOMO1, HOMO2 and HOMO3.}
\begin{ruledtabular}
\begin{tabular}{cccccccc}
& & C$_2$H$_4$ & & CH$_4$ & & C$_3$ H$_8$ & C$_4$H$_{10}$ \\
& & HOMO & HOMO1 & HOMO2 & HOMO3 & HOMO & HOMO \\
\hline
 $I_P$ (eV) & & 10.5 & 12.6 & 12.6 & 12.6 & 10.9 & 10.6 \\
 \hline
 $l$ & $m$ & & & & & \\
 $0$ & $0$ & & & & & & $2.24$\\
 $1$ & $0$ & & & & $-1.65$ & &\\
 $1$ & $\pm 1$ & $-1.09i$ & $\pm 1.16$ & $-1.16 i$& &$\pm 0.36$ &\\
 $2$ & $0$ & & & & & & $-1.87$\\
 $2$ & $\pm 1$ & & $0.36 i$& $\pm 0.36$ && $\mp 0.70$ &\\
 $2$ & $\pm 2$ & & & & $\pm0.36 i$ & &$-3.15 \mp 0.15 i$\\
 $3$ & $0$ & & & & $0.15$ & &\\
 $3$ & $\pm 1$ & $-0.23i$ & $\pm 0.07$ & $-0.07 i$&&$\mp 0.21$ &\\
 $3$ & $\pm 3$ & & $\mp 0.09$ & $-0.09 i$&&$\mp 0.54$ &\\
 $4$ & $0$ & & & & & &$0.99$\\
 $4$ & $\pm 2$ & & & & & &$1.14$\\
 $4$ & $\pm 4$  & & & & & & $1.94$\\
 $4$ & $\pm 3$ & & & &&$\pm 0.20$ &\\
 $5$ & $\pm 3$ & & & & &$\mp 0.06$ &\\
 $6$ & $0$ & & & & & &$-0.28$\\
 $6$ & $\pm 2$ & & & & & &$-0.28$\\
 $6$ & $\pm 4$ & & & & & &$-0.31$\\
 $6$ & $\pm 6$ & & & & & &$-0.46$\\
\end{tabular}
\end{ruledtabular}
\end{table*}

\section{Results and discussion}\label{Results}
In this section we present results on the high-harmonic yield from
ethylene, methane, propane and butane. Since we do not include
effects of propagation~\cite{priori1,tosi1} our calculations cannot
be directly compared with the experimental results of
Ref.~\cite{altucci1}. If our main purpose was the optimization of
the harmonic yield with the object of generating attosecond pulses
then, surely, phase-matching should be taken into account.
Nevertheless, in this theoretical work aiming at isolating and
illustrating clearly the effects of symmetry, degenerate orbitals
and orientation, we find it reasonable to disregard phase-matching:
first, because propagation effects can be reduced experimentally by
using a gas jet which is short compared to the coherence
length~\cite{levesque1}, and second, an understanding of the single
molecular response is needed in order to understand the harmonic
yield from a whole gas of molecules.

In all results presented below, we calculate the signal of harmonics
polarized along a linearly polarized $800$ nm, $1.8\times 10^{14}$
W/cm$^2$ driving laser. As the light is linearly polarized the
results are independent of $\phi$, the rotation around the
polarization vector.
We consider molecules that are either randomly oriented or have been
one- or three-dimensionally oriented. The orientational
distributions $G_{t_d}$ [see Eqs.~\eqref{eq:Gtd}-\eqref{eq:signal}]
corresponding to random, one-dimensional or three-dimensional
orientation are as follows
\begin{eqnarray}
G_{t_d}^\text{random}&=&\frac{1}{8\pi^2},\label{eq:Grandom}\\
G_{t_d}^\text{1D}(\theta',\chi')&=&\frac{1}{4\pi^2}\frac{\delta(\theta'-\theta)}{\sin\theta'},\label{eq:G1D}\\
G_{t_d}^\text{3D}(\theta',\chi')&=&\frac{1}{2\pi}\frac{\delta(\theta'-\theta)}{\sin\theta'}\delta(\chi'-\chi)\label{eq:G3D}.
\end{eqnarray}
Here the BF $z$-axis is oriented at angle $\theta$ with respect to
the LF $Z$-axis in the cases of one- and three-dimensional
orientation and the molecule is rotated an angle $\chi$ around the
BF $z$-axis in the case of three-dimensional orientation.

\subsection{Ethylene: Effects of the orbital symmetry} We first present results on the HHG from ehtylene (C$_2$H$_4$). This molecule
has a non-degenerate HOMO which makes the influence of the HOMO on
the harmonic signal relatively transparent. Additionally, ethylene
is interesting from the point of view that field-free
one-dimensional alignment has been carried out
experimentally~\cite{rouzee2006} and field-free three-dimensional
alignment has been explored theoretically~\cite{underwood2005}.
Consequently, the theoretical results presented below may, in
principle, be subject to experimental investigations.

\begin{figure}
  \begin{center}
    \includegraphics[width=1.0\columnwidth]{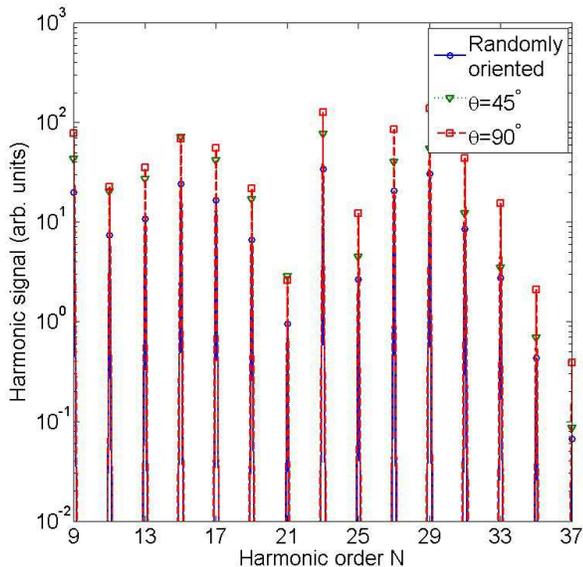}
  \end{center}
  \caption{(Color online) The orientational dependence of the harmonic spectrum from ethylene.
  The absence of even harmonics is explained by the inversion symmetry of the HOMO (see text).}
  \label{fig:fig5}
\end{figure}

Figure~\ref{fig:fig5} shows HHG spectra from ethylene at different
orientations corresponding to the orientational distributions of
Eqs.~\eqref{eq:Grandom} and~\eqref{eq:G1D}. The overall effect of
orienting the molecule is a scaling of the spectrum. The reason for
this scaling is that all harmonics have similar orientational
dependence. We also note the absence of even harmonics in the
spectrum, which is easily explained from the inversion symmetry of
the HOMO of ethylene: The inversion symmetry means that the HOMO is
composed of angular momentum states separated by even multiples of
$\hbar$ (see Table~\ref{tab:table1}). At the same time absorption
and emission of laser photons change the size of the electronic
orbital angular momentum by $\pm\hbar$, so in order both to start
off and end up in one of the angular momentum states of the HOMO an
even number of dipole transitions is necessary. Hence, the HHG
process requires the absorption of an odd number of photons followed
by the emission of a single odd harmonic.

\begin{figure}
  \begin{center}
    \includegraphics[width=0.39\columnwidth]{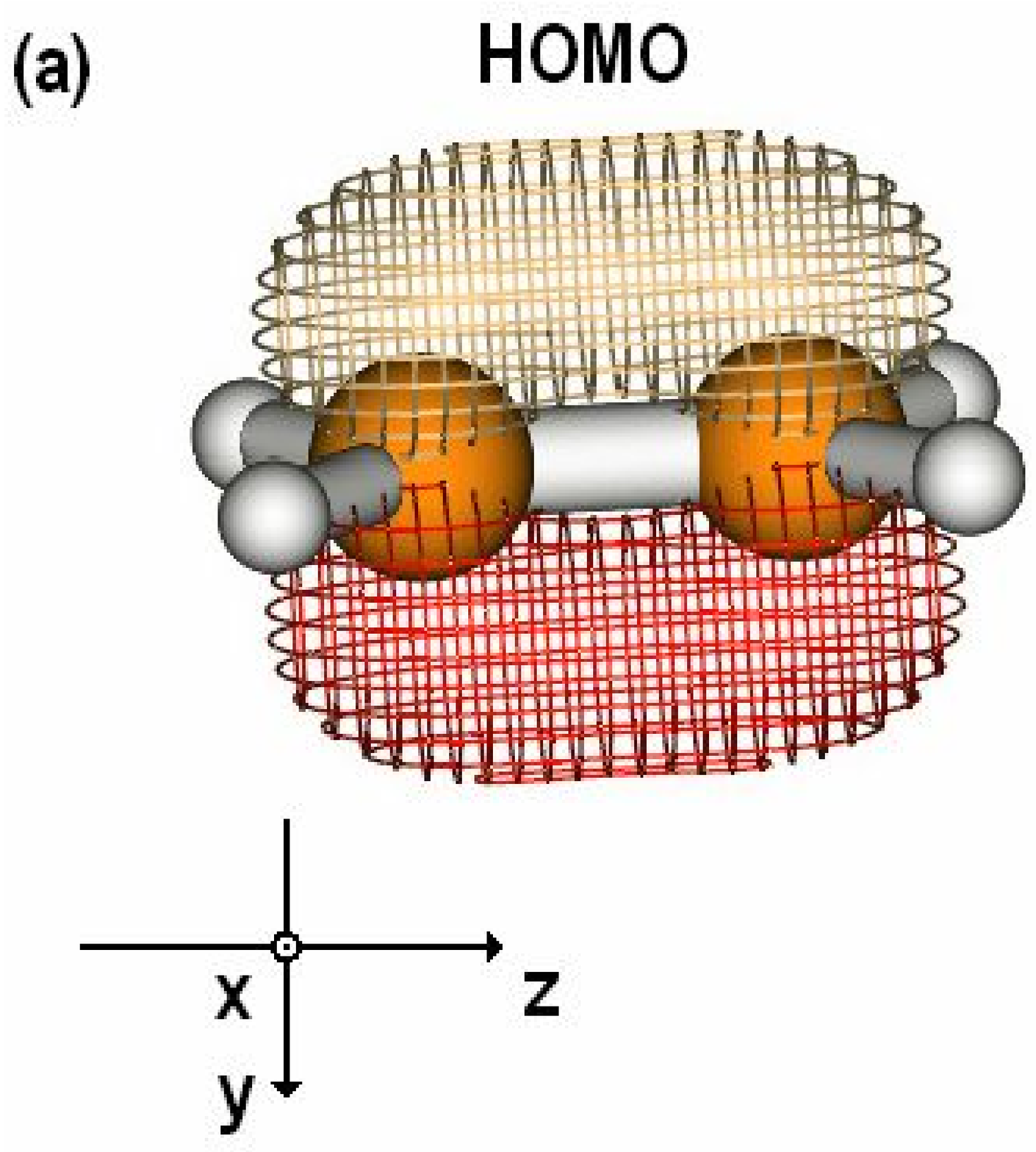}\\
    \includegraphics[width=0.9\columnwidth]{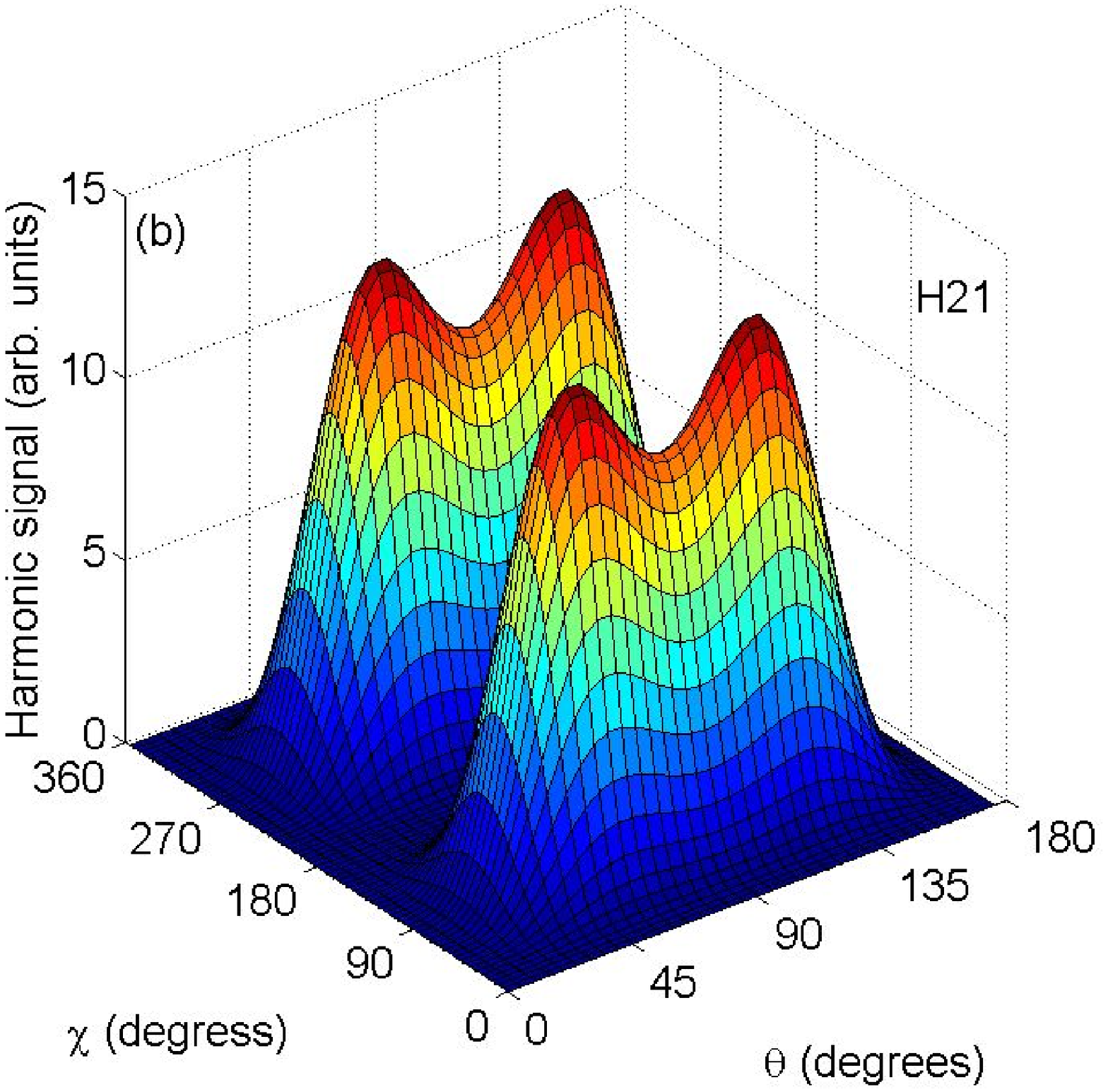}\\
    \includegraphics[width=0.9\columnwidth]{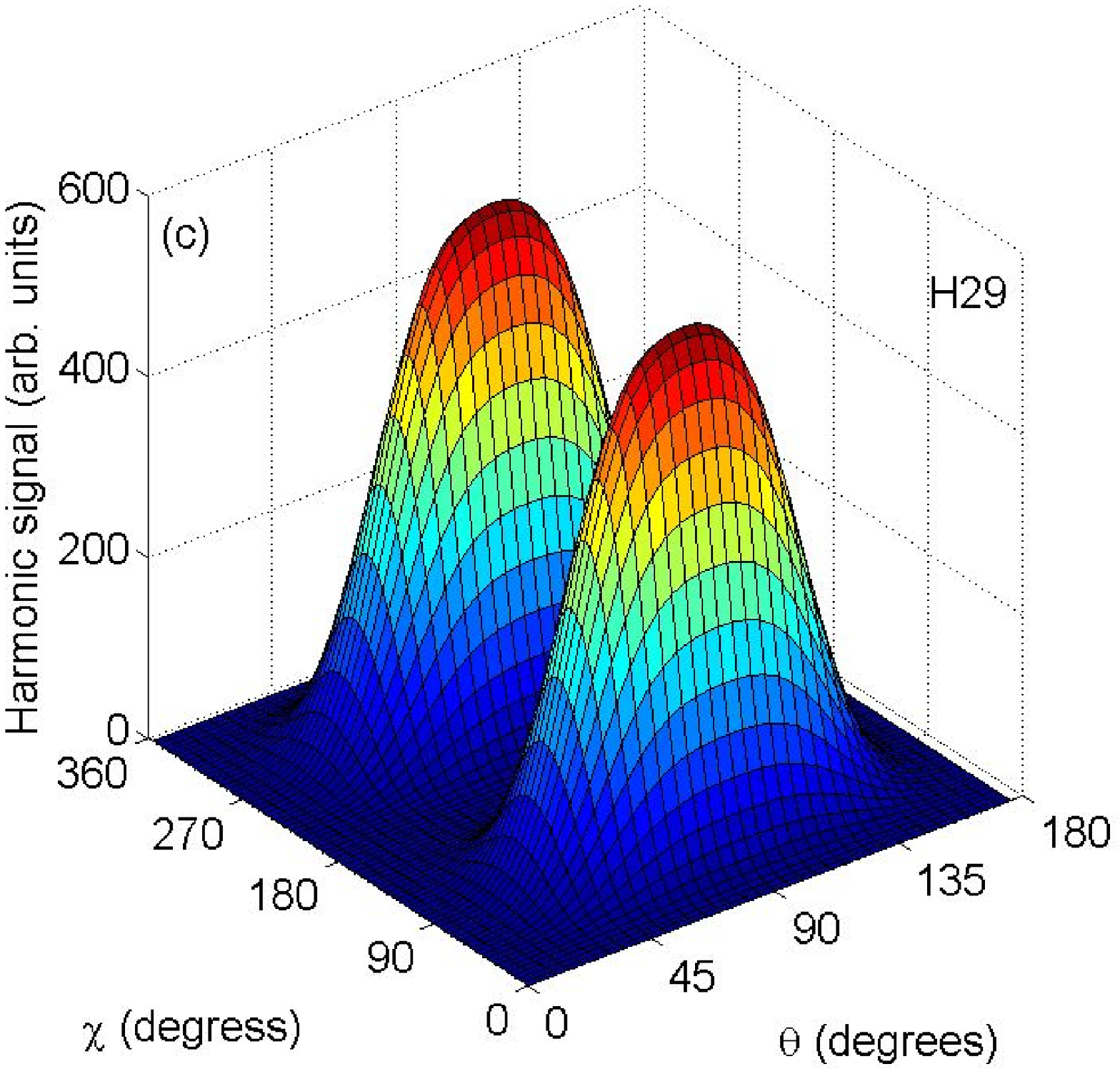}
  \end{center}
  \caption{(Color online) The orientational dependencies of the harmonics reflect the HOMO of ethylene.
  In particular the harmonics vanish, when the laser polarization
  coincides with the nodal plane of the HOMO.
  Panel (a) shows the geometry of ethylene (C$_2$H$_4$) along with an isocontour for the
  HOMO. Dark shading (red online) denotes a negative sign of the HOMO wave function. Light shading (brownish online) denotes a positive sign. Directions of the BF axes are shown. We always choose the center of mass as the origin
  of the BF coordinate system.
  Panels (b) and (c) present the dependencies of the $21$st (H21) and $29$th (H29) harmonics on orientation as given by Euler angles $\theta$ and $\chi$.}
  \label{fig:fig6}
\end{figure}

We have investigated HHG from C$_2$H$_4$ that has been fixed in both
$\theta$- and $\chi$-angles corresponding to the three-dimensional
orientation given by Eq.~\eqref{eq:G3D}. Figure~\ref{fig:fig6} shows
representative results of the calculations. In order to understand
the results, we also show the HOMO of ethylene in the figure.
The meaning of the Euler angles was explained below
Eq.~\eqref{eq:HOMO}. In the model used to simulate HHG an electron
has to escape along the laser polarization axis [cf.
Eq.~\eqref{Aamp}]. This is impossible, if the polarization axis lies
along the nodal plane, which is the reason for the vanishing
harmonic signal, when either $\theta=0^\circ$ $(180^\circ)$ or
$\chi=0^\circ$ $(180^\circ,360^\circ)$. When the molecule is rotated
the nodal plane is removed from the polarization axis of the laser
and the strength of the harmonics increases. As seen from
Figs.~\ref{fig:fig6}(b) and (c) the harmonics peak at different
values of the Euler angle $\theta$. The varying positions of the
peaks arise from different orientational behavior of the ionization
and propagation-recombination steps making up the HHG process [cf.
the discussion below Eq.~\eqref{harmonic}-\eqref{Bamp}]: As the
electron escapes along the polarization direction the ionization is
maximal when $\theta$ lies in between $0^\circ$ and
$90^\circ$~\cite{kjeldsen1}. The propagation-recombination step,
however, is optimized when $\theta=90^\circ$, but the width of the
peak depends on the harmonic order. These observations account for
the different orientational behavior of the harmonics shown on
Figs.~\ref{fig:fig6}(b) and (c).

We note, in passing, that a set of data as the ones presented in
Figs.~\ref{fig:fig6}(b) and (c) for a full range of harmonic
energies would, in principle, allow a tomographic reconstruction of
the HOMO~\cite{itatani04}.

\subsection{Methane: Interference of degenerate HOMOs}
\begin{figure}
  \begin{center}
    \includegraphics[width=0.5\columnwidth]{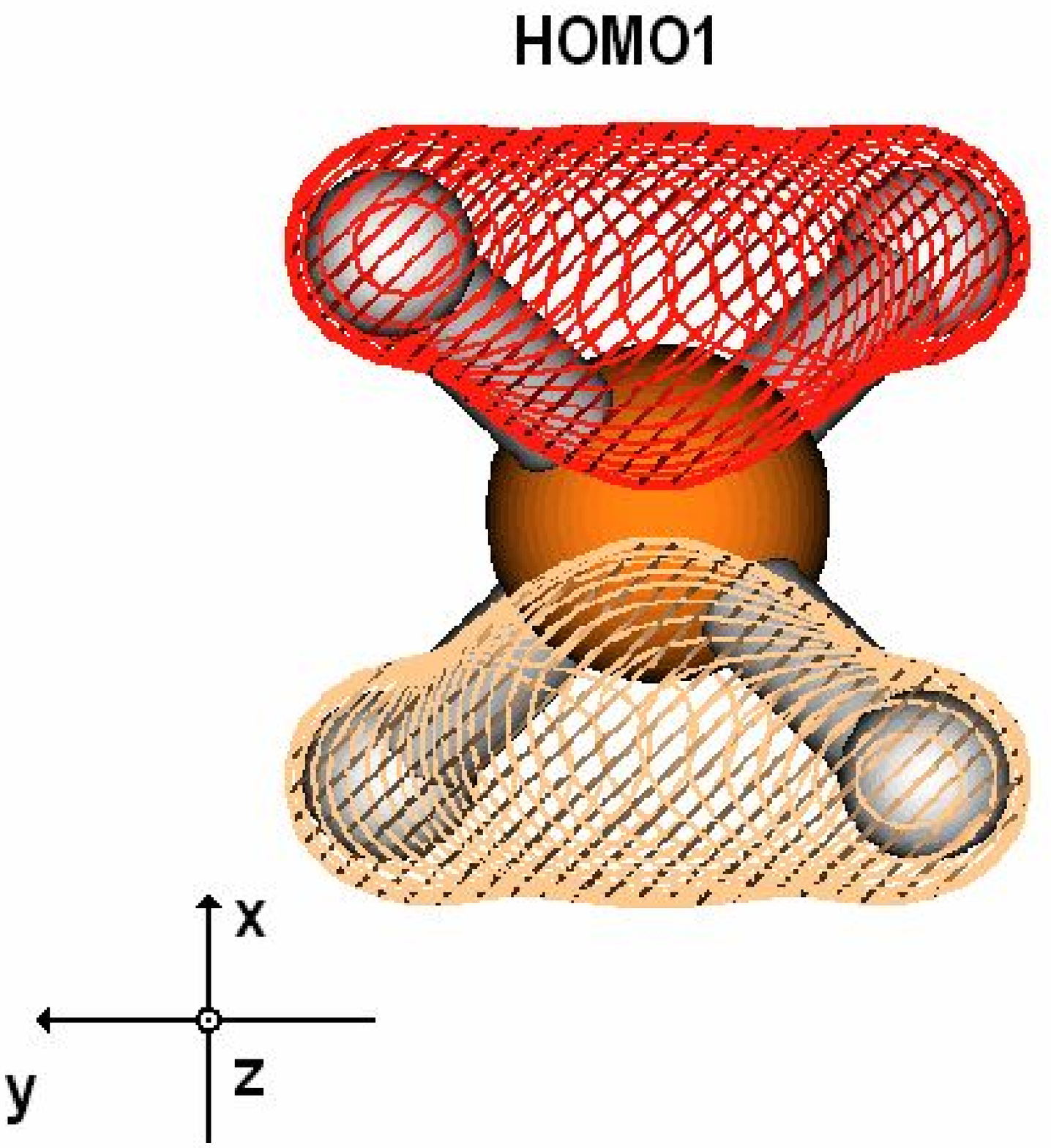}\\
    \includegraphics[width=0.3\columnwidth]{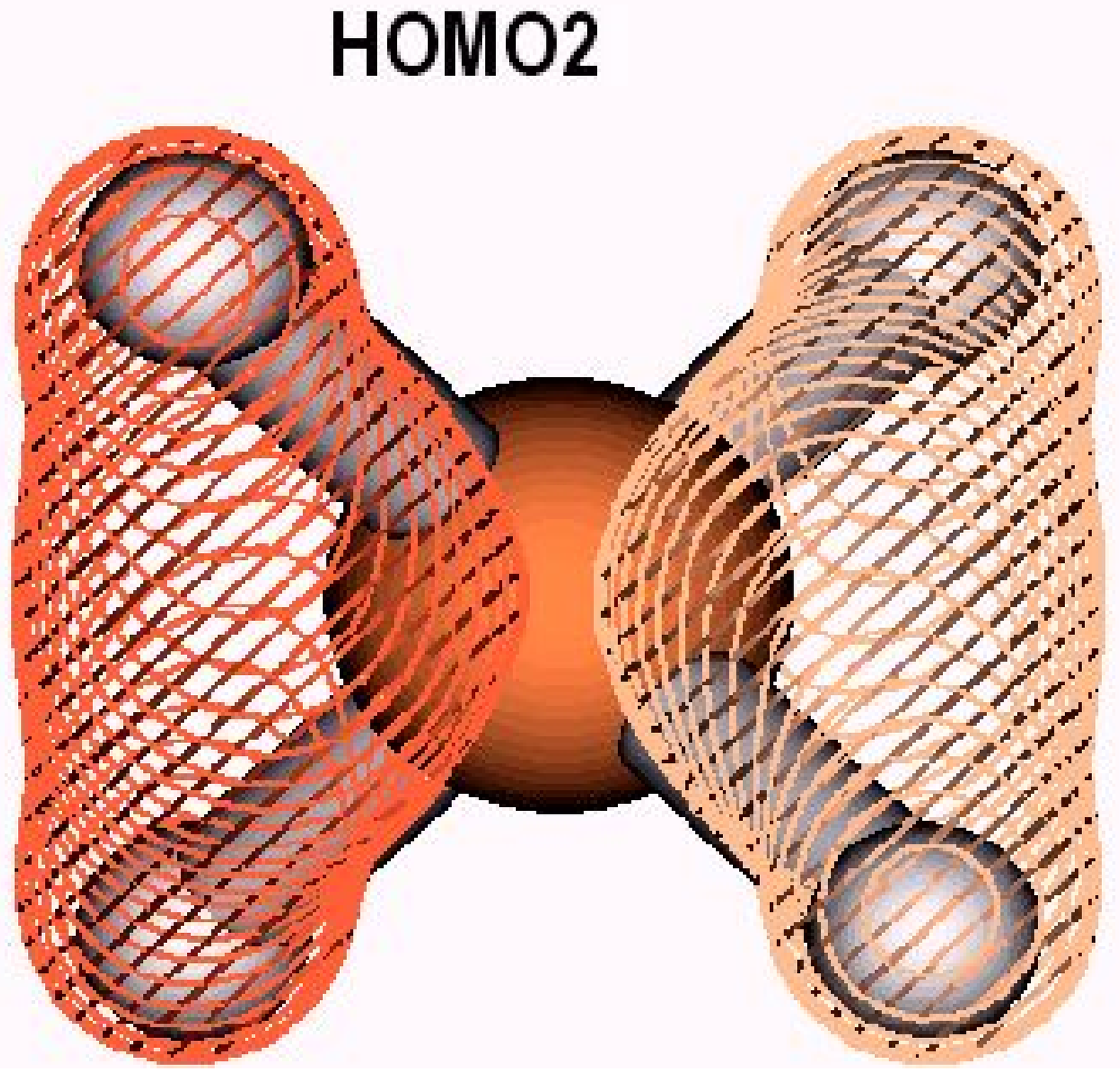}
    \includegraphics[width=0.3\columnwidth]{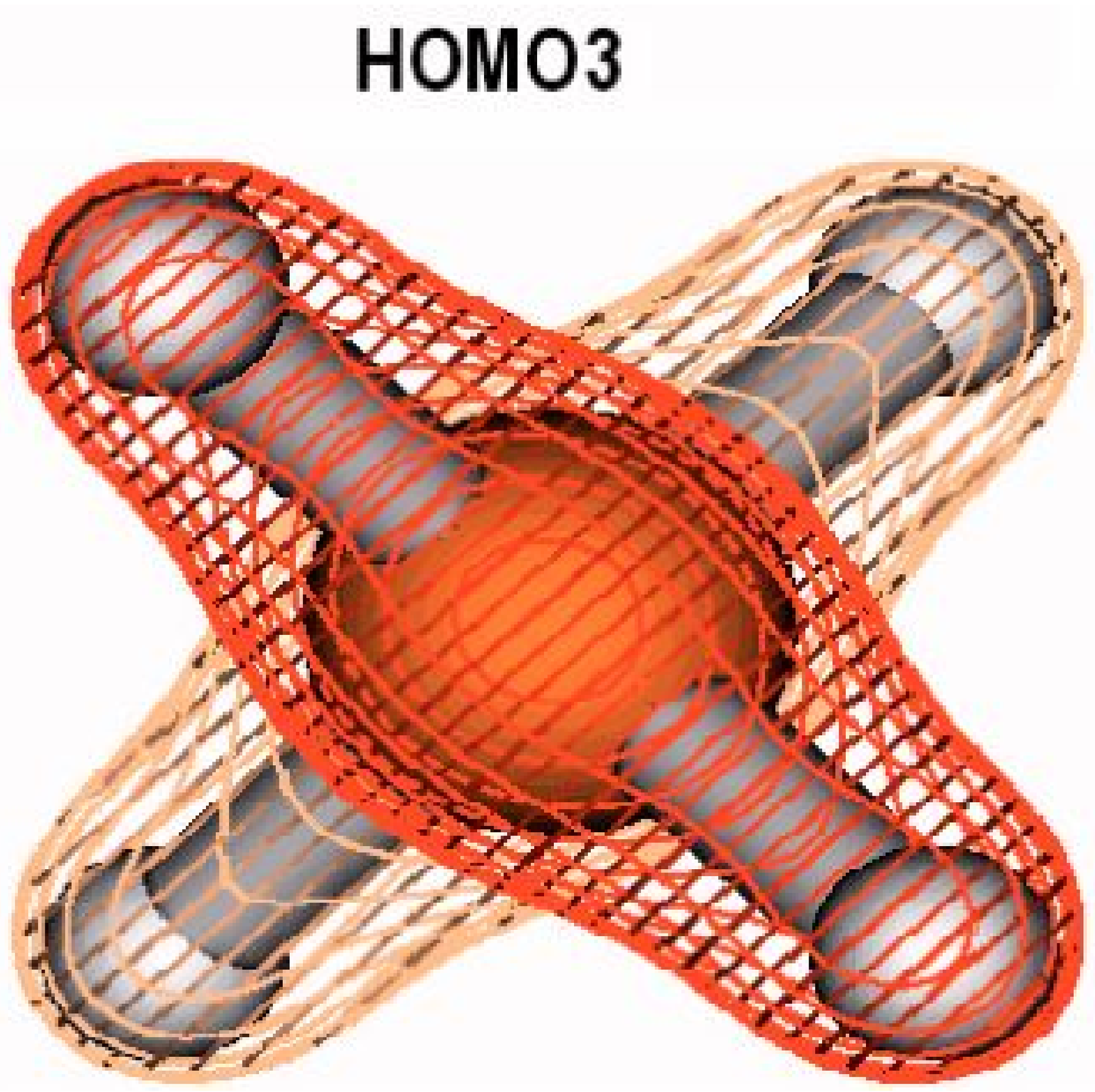}
  \end{center}
  \caption{(Color online) The geometry of the methane molecule (CH$_4$) along
  with isocontours of the three degenerate HOMOs. The signs of the HOMO wave
  functions are indicated by the coloring, where dark shading (red online)
  denotes a negative sign and a light shading (brownish online) denotes a
  positive sign. The coordinate system shows the directions of the BF axes.
  Note that the HOMOs differ from one another only by a simple rotation.
}
  \label{fig:fig1}
\end{figure}

\begin{figure}
  \begin{center}
    \includegraphics[width=1.0\columnwidth]{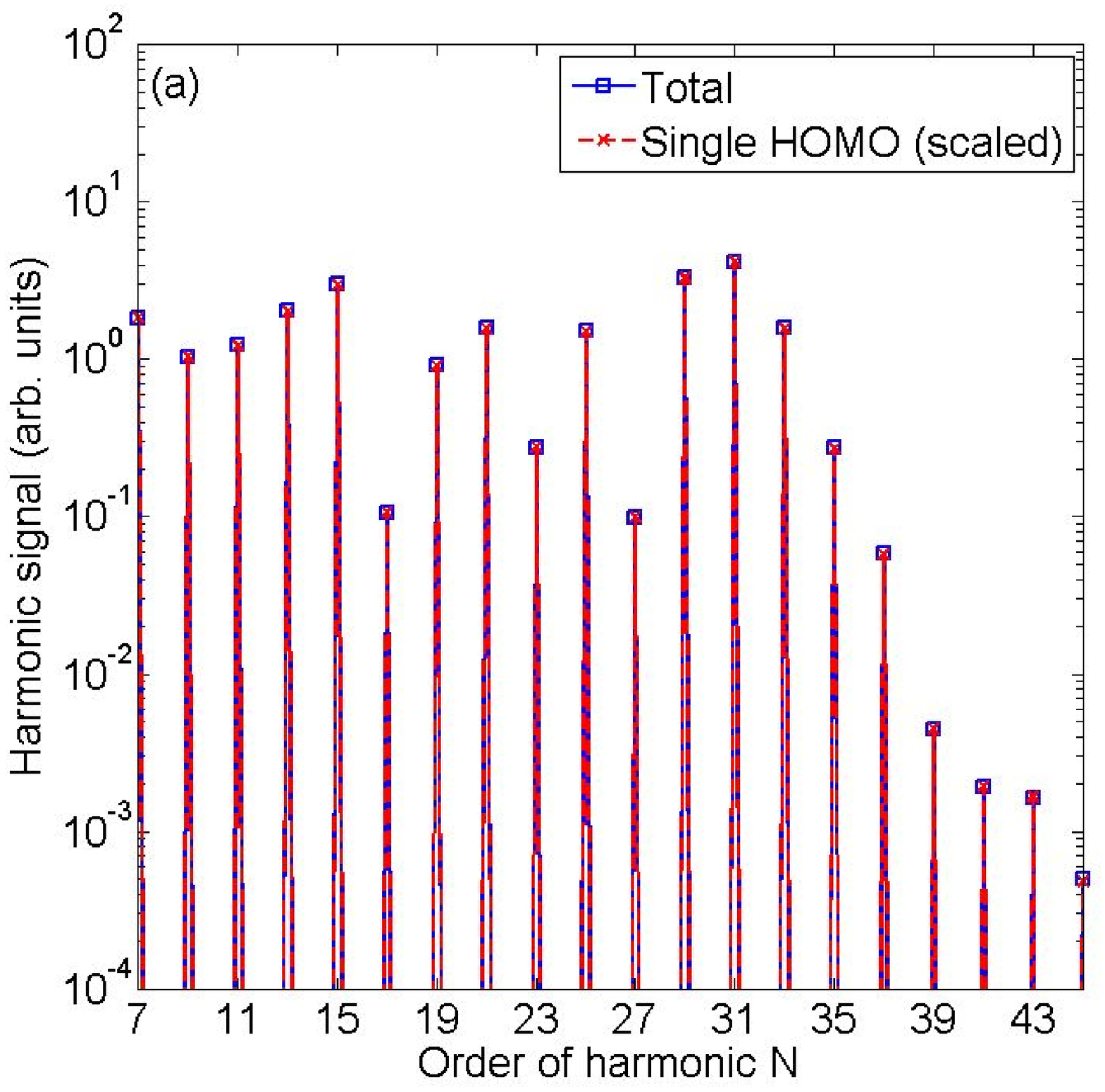}\\
    \includegraphics[width=1.0\columnwidth]{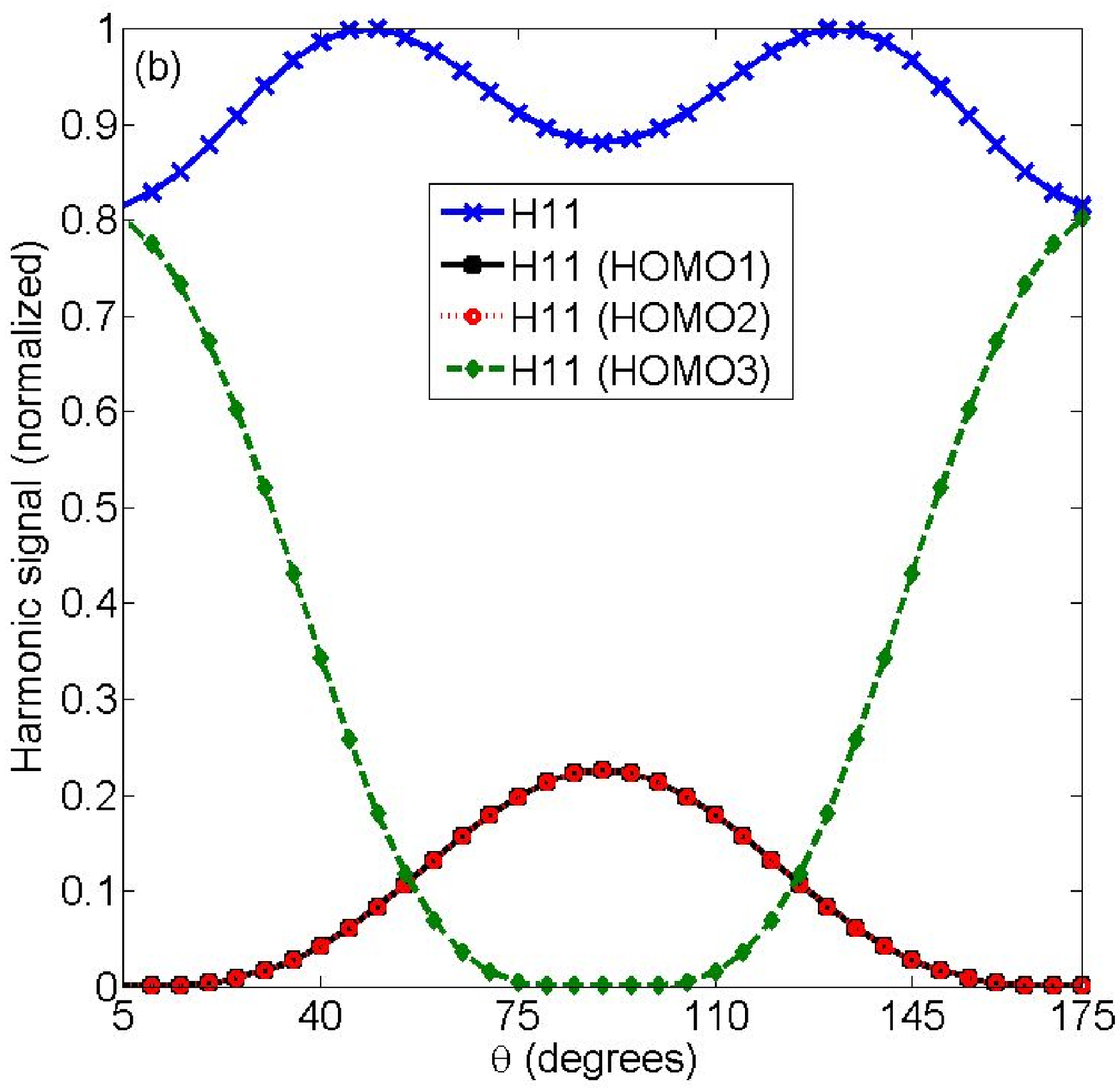}
  \end{center}
  \caption{(Color online) The figure illustrates the interference of high-harmonics coming from different
  orbitals as prescribed by Eq.~\eqref{eq:signal}. The physical system used is methane (CH$_4$). Panel (a) shows the total harmonic yield from
  randomly oriented methane. It is just a multiple of the harmonic yield from a
single HOMO. Even harmonics vanish as a result of the orientational
averaging. Panel (b) illustrates the detailed
  $\theta$-dependence of the $11$th harmonic (H11). Although not measurable, we also show the single
  HOMO signals. The calculated signals have all been normalized to the
  maximum total yield of the $11$th harmonic. The figure underlines the importance of adding coherently the amplitude contributions from
  each HOMO, $\lambda=1,2,3$.}
  \label{fig:fig2}
\end{figure}

We now turn to the harmonic yield from the methane molecule
(CH$_4$). Methane has three degenerate HOMOs as shown on
Fig.~\ref{fig:fig1}, and we can use this molecule to demonstrate the
effect discussed below Eq.~\eqref{eq:signal}, i.e., the absence of
interference effects from randomly oriented molecules, when
degenerate HOMOs differ only by a rotation. To this end we have
compared the total harmonic yield with the yield from a single HOMO
when the orientational distribution is as given by
Eq.~\eqref{eq:Grandom} and confirmed that the results agree except
from an overall scaling factor. This is illustrated on
Fig~\ref{fig:fig2}(a).
%

Next, we have carried out calculations of HHG from methane that has
been one-dimensionally oriented with the BF $z$-axis at some fixed
angle $\theta$ relative to the polarization direction. The
orientational distribution used is given by Eq.~\eqref{eq:G1D}. We
do not show the harmonic spectrum in this case, since it does not
differ much in structure from Fig.~\ref{fig:fig2}(a). This is
probably due to the fact that methane is rather small and compact,
which makes it spherical-like after $\chi$-averaging. Consequently,
no structure is revealed by the electrons, not even the most
energetic, and the harmonics exhibit the same overall
$\theta$-dependence.

On Fig.~\ref{fig:fig2}(b) we show this typical angular dependence of
the harmonics. In the figure the upper curve shows the signal when
the coherence between the individual HOMOs is correctly accounted
for [cf. Eq.~\eqref{eq:signal}]. The other curves in the figure show
the unphysical signals from each HOMO. Clearly, this figure
illustrates that there is a strong interference between the single
HOMO
amplitudes in the angle resolved signal. 
We may understand the single HOMO signals in Fig.~\ref{fig:fig2}(b)
from the structure of the HOMOs: First, we explain the dips. The
vanishing signals of HOMO1 and HOMO2 at $\theta=0^\circ$ are
explained by the fact that in this case the polarization vector
points along the BF $z$-axis, hence coincinding with the nodal
planes of these HOMOs, as seen from Fig.~\ref{fig:fig1}.
Furthermore, the electron causing HHG must escape along the
polarization axis to Eq.~\eqref{Aamp}, and we conclude that these
two HOMOs cannot generate harmonics for $\theta=0^\circ$. At
$\theta=90^\circ$ the polarization axis is directed into the BF
$xy$-plane which is the nodal plane of HOMO3 (see
Fig.~\ref{fig:fig1}) and consequently HHG from HOMO3 is excluded.
Second, we remark that the single HOMO-yield at a given value of
$\theta$ is obtained by averaging amplitudes from all degrees of
rotation around the BF $z$ [cf.-axis~\eqref{eq:signal}
and~\eqref{eq:G1D}]. It is therefore obvious from symmetry that the
results of HOMO1 and HOMO2 must be identical.

%

\subsection{Propane and butane: Effects of orientation}
\begin{figure}
  \begin{center}
    \includegraphics[width=0.5\columnwidth]{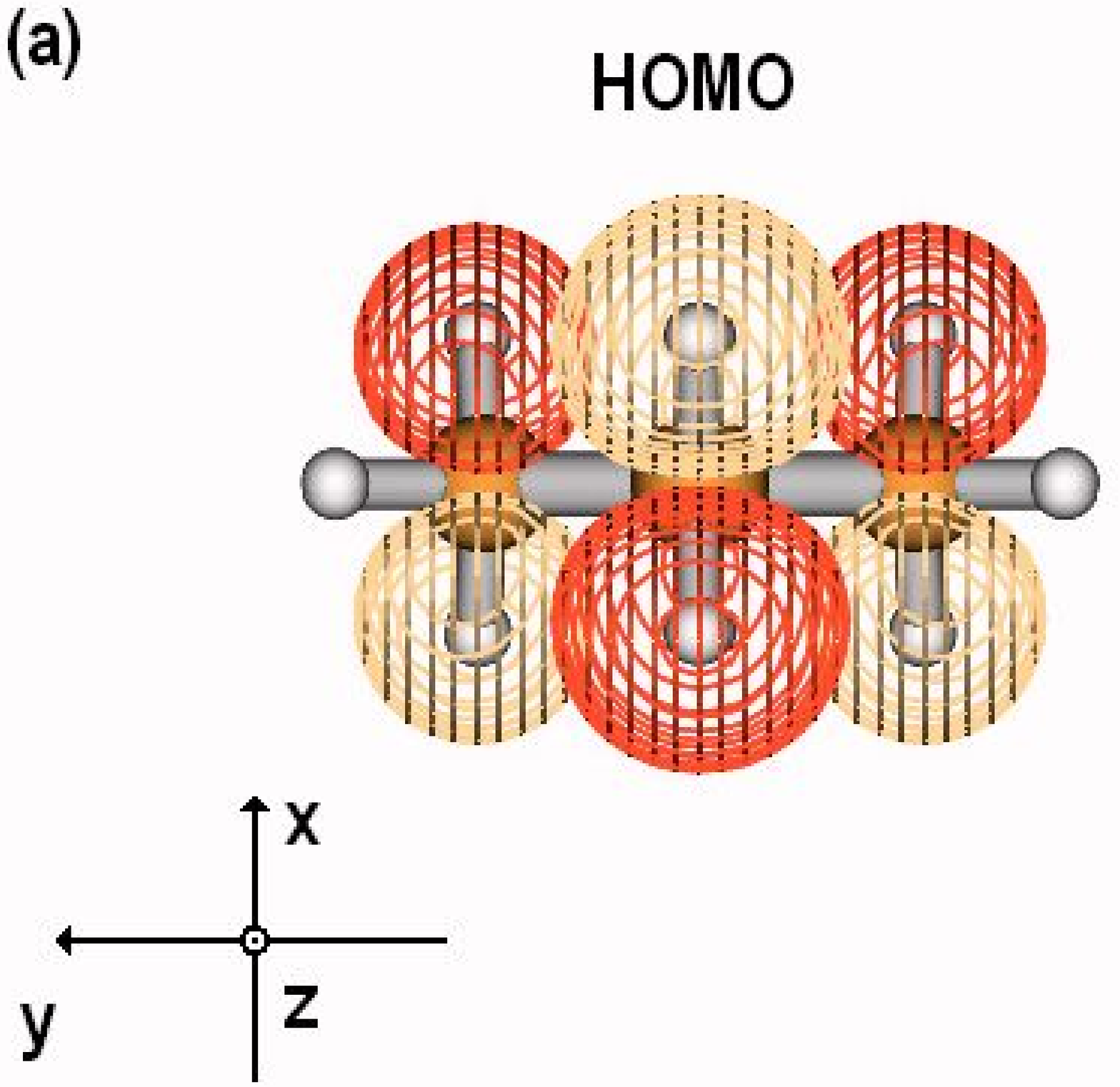}\\
    \includegraphics[width=1.0\columnwidth]{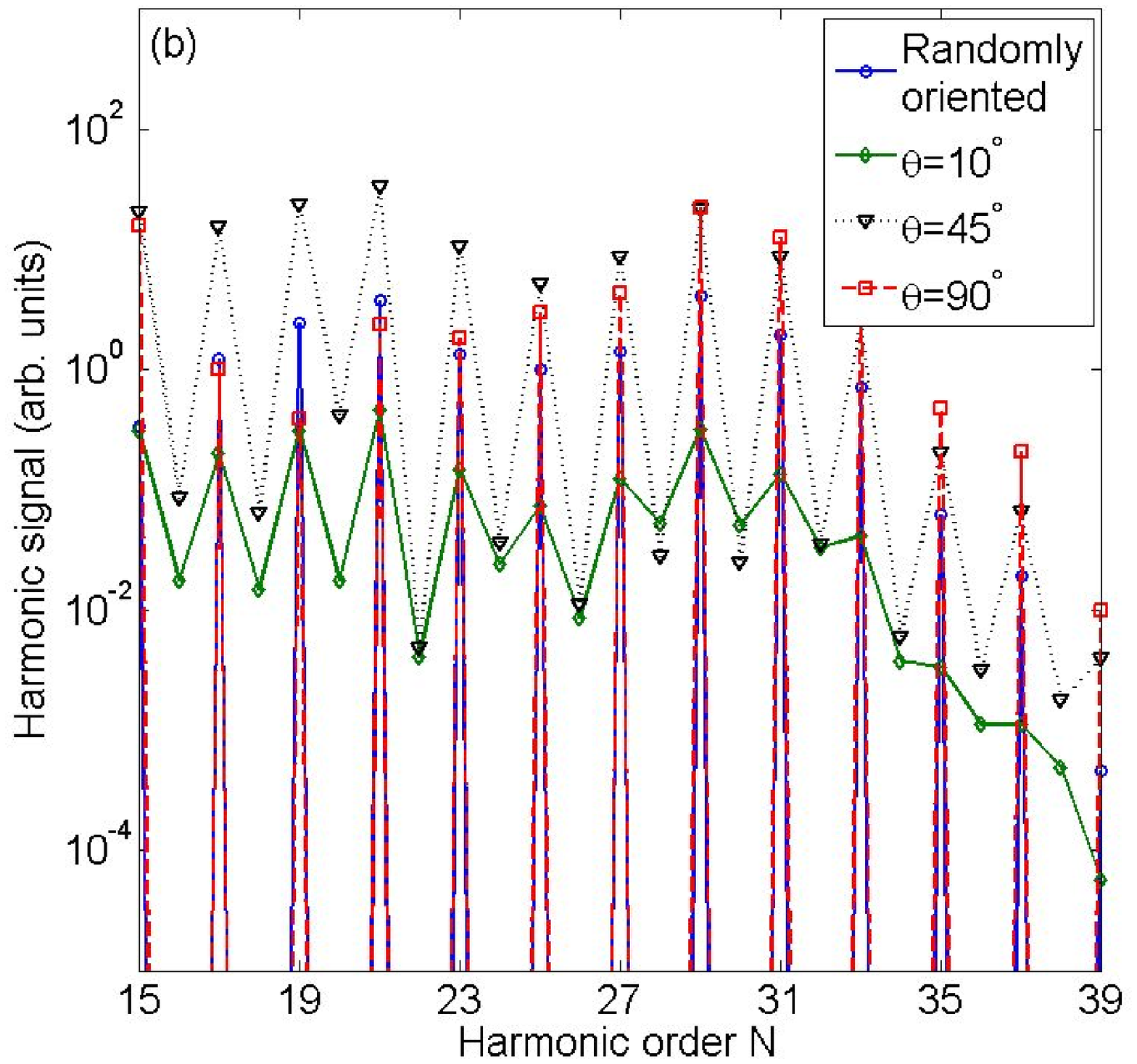}
  \end{center}
  \caption{(Color online) The figure clearly illustrates how the open and extended structure
  of propane leads to a harmonic spectrum with a complicated orientational dependence.
  Panel (a) shows the geometry of the propane molecule (C$_3$H$_8$) along with an isocontour of the
  HOMO. Dark shading (red online) denotes a negative sign of the
  HOMO wave function and light shading (brownish online) denotes a
  positive sign. The $xyz$-axes of the BF frame is shown. Panel (b) illustrates the dependence of the harmonic spectrum on
  the orientation of the propane molecule. At the orientation $\theta=90^\circ$ even harmonics
  are suppressed as discussed in the text.}
  \label{fig:fig3}
\end{figure}

In the following we consider propane (C$_3$H$_8$) and butane
(C$_4$H$_{10}$). We begin with numerical results on the propane
molecule. Figure~\ref{fig:fig3} shows the HOMO of propane along with
the harmonic signal in randomly oriented and one-dimensionally
oriented scenarios corresponding to orientational distributions from
Eqs.~\eqref{eq:Grandom} and~\eqref{eq:G1D}.
The dependence of the HHG spectrum on orientation amounts to more
than just a simple scaling, e.g., the $19$th harmonic is alternately
above and below the neighboring odd harmonics depending on the
orientation of the propane molecule. We ascribe this to the fact
that propane is a rather extended and open-structured molecule. In
general, when the BF $z$-axis is close to the LF $Z$-axis the HHG is
suppressed, because the almost vanishing wave function along the
polarization direction [see Fig.~\ref{fig:fig3}(a)] makes the ATI
amplitudes in Eq.~\eqref{harmonic} small. Notice also that the even
harmonics disappear at an orientation of $90^\circ$. This is due to
the fact that every photon absorption at this orientation changes
the projection of the electron angular momentum on the $Z$-axis with
$\pm\hbar$, and it is seen from Table~\ref{tab:table1} that an odd
number of photon absorptions is necessary in order to start off and
end up in the HOMO.

We remark that harmonics of order $15$ to $27$ are suppressed for
$\theta=90^\circ$, which means that alignment is a tool for
favorising the harmonics close to and above the semiclassical cutoff
energy at $I_p+3.17U_p\simeq29\times\hbar\omega$. These harmonics
are known to be well phase-synchronized in the case of diatomic
molecules~\cite{lu2006}, and this may also hold true in the case of
propane. Furthermore, phase-synchronized harmonics are
synchronically emitted and the superposition of such harmonics
constitute the basis for the generation of attosecond pulses.

\begin{figure}
  \begin{center}
    \includegraphics[width=0.5\columnwidth]{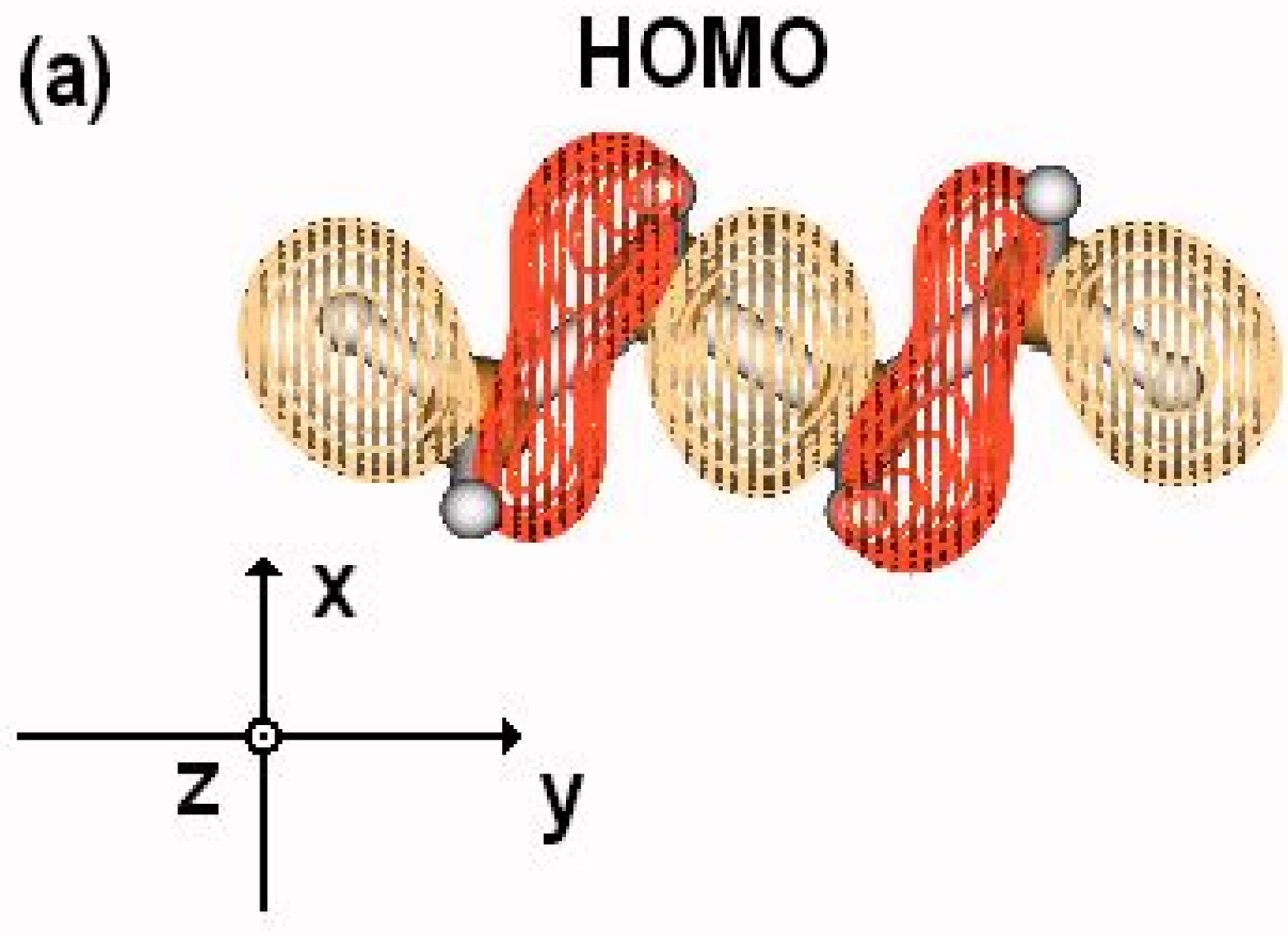}\\
    \includegraphics[width=1.0\columnwidth]{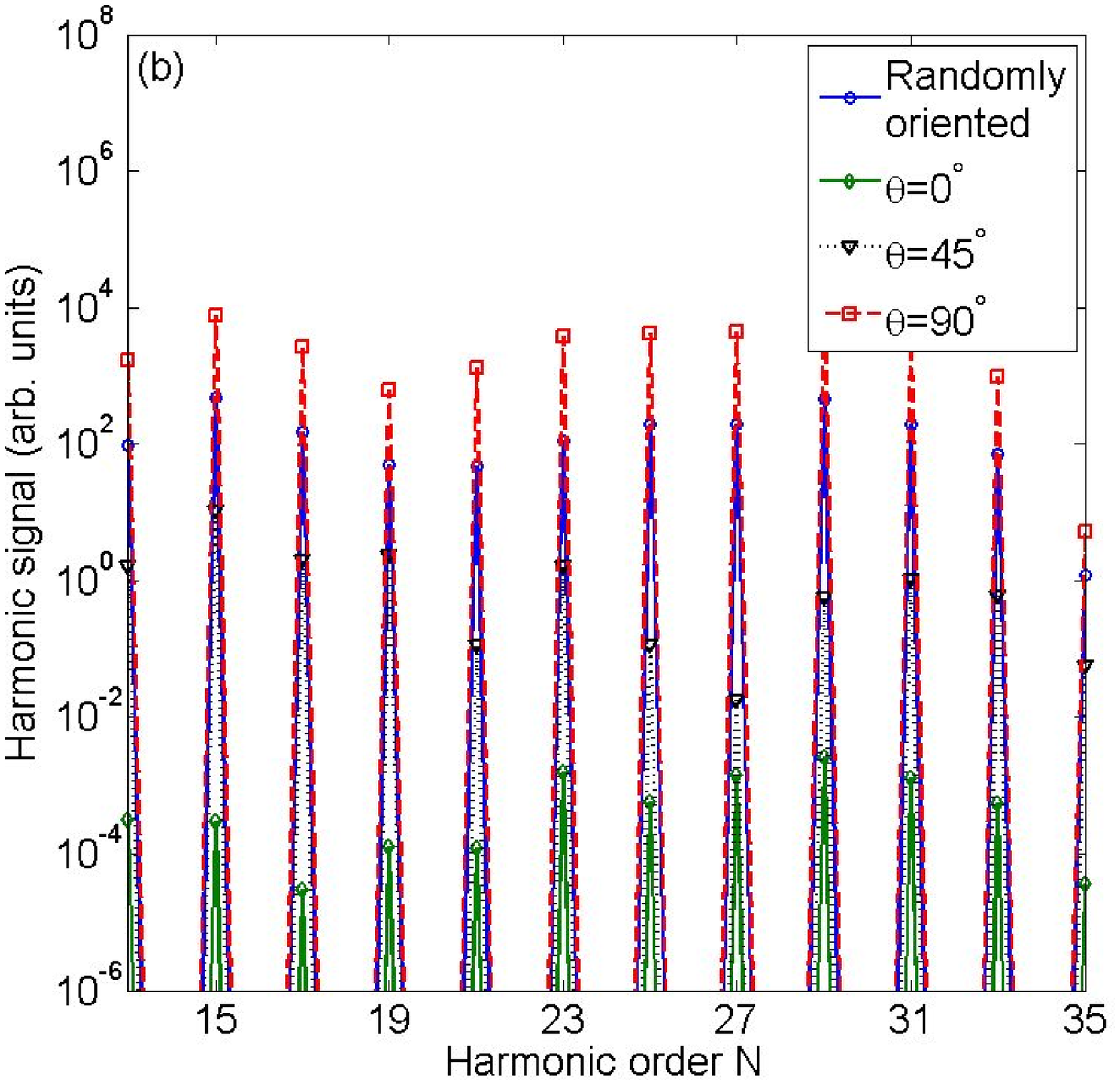}
  \end{center}
  \caption{(Color online) Panel (a) shows the geometry of the butane molecule (C$_4$H$_{10}$) and an isocontour of the HOMO.
  The sign of the HOMO wave function is indicated by the coloring, where dark
  shading (red online) denotes a negative sign and light shading
  (brownish online) denotes a positive sign.
  We also show the coordinate system of the BF frame. Panel (b) shows the harmonic spectrum corresponding to different orientations.
  As in the case of ethylene the HOMO is inversion symmetric which excludes the presence of even harmonics.}
  \label{fig:fig4}
\end{figure}

Figure~\ref{fig:fig4} illustrates the HHG from butane.
As with propane we observe a complicated behavior of the individual
harmonics upon the orientation. Opposite to the case of propane,
there is no particular orientation that favors harmonics near the
semiclassical cutoff energy.

\section{Conclusions and outlook}\label{Conclusion}
In the present work we have discussed general issues related to the
effects of molecular symmetry, degeneracy and orientation in HHG. To
this end, we have considered the high-harmonic signal from several
polyatomic molecules, namely ethylene, methane, propane and butane.
In the case of ethylene we have shown how the dependence of
orientation reflects the HOMO: The strength of the harmonics is
increased when the laser polarization is directed away from the
nodal plane, but the detailed orientational dependence differ from
one harmonic to another. We have used methane as an example of a
molecule with degenerate HOMOs. Then harmonic amplitudes of
different HOMOs need to be added coherently, and interference
effects are in general unavoidable. Although, some information about
the individual HOMOs can be extracted by orienting the molecule we
illustrated the importance of including all HOMOs when calculating
the harmonic signal from one-dimensionally oriented methane.
Finally, the propane and butane molecules served as illustrations of
HHG from extended structures, and the individual harmonics carry
their own characteristic orientational dependence.
In the case of propane harmonics near the cutoff can be promoted by
means of orientation and a selection of these energetic harmonics is
of interest to attosecond pulse generation.

While we expect the conclusions drawn above to be fairly model
independent we would like to mention a few open theoretical
questions that should be addressed in the near future. In the
present paper a relatively simple model using several
approximations, including the strong-field approximation and a
saddle-point approximation, has been applied to describe HHG. The
model is formulated in length gauge, but at this level of
approximation it is still to be settled wether another gauge would
be more appropriate~\cite{lein2006,chen2007}. In addition, a future
challenge is to take into account the finite duration of the
harmonic driving laserpulse. This is necessary in order to
investigate further the possibility of using large molecules as
sources of attosecond pulses.

\begin{acknowledgments}
The present work was supported by the Danish Research Agency (Grant.
No. 2117-05-0081).
\end{acknowledgments}


\begin{thebibliography}{26}
\expandafter\ifx\csname
natexlab\endcsname\relax\def\natexlab#1{#1}\fi
\expandafter\ifx\csname bibnamefont\endcsname\relax
  \def\bibnamefont#1{#1}\fi
\expandafter\ifx\csname bibfnamefont\endcsname\relax
  \def\bibfnamefont#1{#1}\fi
\expandafter\ifx\csname citenamefont\endcsname\relax
  \def\citenamefont#1{#1}\fi
\expandafter\ifx\csname url\endcsname\relax
  \def\url#1{\texttt{#1}}\fi
\expandafter\ifx\csname urlprefix\endcsname\relax\def\urlprefix{URL
}\fi \providecommand{\bibinfo}[2]{#2}
\providecommand{\eprint}[2][]{\url{#2}}

\bibitem[{\citenamefont{Sansone et~al.}(2006)\citenamefont{Sansone, Benedetti,
  Calegari, Vozzi, Avaldi, Flammini, Poletto, Villoresi, Altucci, Velotta
  et~al.}}]{Sansone06}
\bibinfo{author}{\bibfnamefont{G.}~\bibnamefont{Sansone}},
  \bibinfo{author}{\bibfnamefont{E.}~\bibnamefont{Benedetti}},
  \bibinfo{author}{\bibfnamefont{F.}~\bibnamefont{Calegari}},
  \bibinfo{author}{\bibfnamefont{C.}~\bibnamefont{Vozzi}},
  \bibinfo{author}{\bibfnamefont{L.}~\bibnamefont{Avaldi}},
  \bibinfo{author}{\bibfnamefont{R.}~\bibnamefont{Flammini}},
  \bibinfo{author}{\bibfnamefont{L.}~\bibnamefont{Poletto}},
  \bibinfo{author}{\bibfnamefont{P.}~\bibnamefont{Villoresi}},
  \bibinfo{author}{\bibfnamefont{C.}~\bibnamefont{Altucci}},
  \bibinfo{author}{\bibfnamefont{R.}~\bibnamefont{Velotta}},
  \bibnamefont{et~al.}, \bibinfo{journal}{Science}
  \textbf{\bibinfo{volume}{314}}, \bibinfo{pages}{443} (\bibinfo{year}{2006}).

\bibitem[{\citenamefont{Carrera et~al.}(2006)\citenamefont{Carrera, Tong, and
  Chu}}]{carrera06}
\bibinfo{author}{\bibfnamefont{J.~J.} \bibnamefont{Carrera}},
  \bibinfo{author}{\bibfnamefont{X.~M.} \bibnamefont{Tong}}, \bibnamefont{and}
  \bibinfo{author}{\bibfnamefont{S.-I.} \bibnamefont{Chu}},
  \bibinfo{journal}{Phys. Rev. A} \textbf{\bibinfo{volume}{74}},
  \bibinfo{pages}{023404} (\bibinfo{year}{2006}).

\bibitem[{\citenamefont{Cao et~al.}(2007)\citenamefont{Cao, Lu, Lan, Hong, and
  Wang}}]{cao07}
\bibinfo{author}{\bibfnamefont{W.}~\bibnamefont{Cao}},
  \bibinfo{author}{\bibfnamefont{P.}~\bibnamefont{Lu}},
  \bibinfo{author}{\bibfnamefont{P.}~\bibnamefont{Lan}},
  \bibinfo{author}{\bibfnamefont{W.}~\bibnamefont{Hong}}, \bibnamefont{and}
  \bibinfo{author}{\bibfnamefont{X.}~\bibnamefont{Wang}}, \bibinfo{journal}{J.
  Phys. B} \textbf{\bibinfo{volume}{40}}, \bibinfo{pages}{869}
  (\bibinfo{year}{2007}).

\bibitem[{\citenamefont{Haworth et~al.}(2007)\citenamefont{Haworth,
  Chipperfield, Robinson, Knight, Marangos, and Tisch}}]{haworth07}
\bibinfo{author}{\bibfnamefont{C.~A.} \bibnamefont{Haworth}},
  \bibinfo{author}{\bibfnamefont{L.~E.} \bibnamefont{Chipperfield}},
  \bibinfo{author}{\bibfnamefont{J.~S.} \bibnamefont{Robinson}},
  \bibinfo{author}{\bibfnamefont{P.~L.} \bibnamefont{Knight}},
  \bibinfo{author}{\bibfnamefont{J.~P.} \bibnamefont{Marangos}},
  \bibnamefont{and} \bibinfo{author}{\bibfnamefont{J.~W.~G.}
  \bibnamefont{Tisch}}, \bibinfo{journal}{Nature Physics}
  \textbf{\bibinfo{volume}{3}}, \bibinfo{pages}{52} (\bibinfo{year}{2007}).

\bibitem[{\citenamefont{Itatani et~al.}(2004)\citenamefont{Itatani, Levesque,
  Zeidler, Niikura, P\'{e}pin, Kieffer, Corkum, and Villeneuve}}]{itatani04}
\bibinfo{author}{\bibfnamefont{J.}~\bibnamefont{Itatani}},
  \bibinfo{author}{\bibfnamefont{J.}~\bibnamefont{Levesque}},
  \bibinfo{author}{\bibfnamefont{D.}~\bibnamefont{Zeidler}},
  \bibinfo{author}{\bibfnamefont{H.}~\bibnamefont{Niikura}},
  \bibinfo{author}{\bibfnamefont{H.}~\bibnamefont{P\'{e}pin}},
  \bibinfo{author}{\bibfnamefont{J.~C.} \bibnamefont{Kieffer}},
  \bibinfo{author}{\bibfnamefont{P.~B.} \bibnamefont{Corkum}},
  \bibnamefont{and} \bibinfo{author}{\bibfnamefont{D.~M.}
  \bibnamefont{Villeneuve}}, \bibinfo{journal}{Nature (London)}
  \textbf{\bibinfo{volume}{432}}, \bibinfo{pages}{867} (\bibinfo{year}{2004}).

\bibitem[{\citenamefont{Torres et~al.}(2007)\citenamefont{Torres, Kajumba,
  Underwood, Robinson, Baker, Tisch, de~Nalda, Bryan, Velotta, Altucci
  et~al.}}]{torres1}
\bibinfo{author}{\bibfnamefont{R.}~\bibnamefont{Torres}},
  \bibinfo{author}{\bibfnamefont{N.}~\bibnamefont{Kajumba}},
  \bibinfo{author}{\bibfnamefont{J.~G.} \bibnamefont{Underwood}},
  \bibinfo{author}{\bibfnamefont{J.~S.} \bibnamefont{Robinson}},
  \bibinfo{author}{\bibfnamefont{S.}~\bibnamefont{Baker}},
  \bibinfo{author}{\bibfnamefont{J.~W.~G.} \bibnamefont{Tisch}},
  \bibinfo{author}{\bibfnamefont{R.}~\bibnamefont{de~Nalda}},
  \bibinfo{author}{\bibfnamefont{W.~A.} \bibnamefont{Bryan}},
  \bibinfo{author}{\bibfnamefont{R.}~\bibnamefont{Velotta}},
  \bibinfo{author}{\bibfnamefont{C.}~\bibnamefont{Altucci}},
  \bibnamefont{et~al.}, \bibinfo{journal}{Phys. Rev. Lett.}
  \textbf{\bibinfo{volume}{98}}, \bibinfo{pages}{203007}
  (\bibinfo{year}{2007}).

\bibitem[{\citenamefont{Patchkovskii et~al.}(2007)\citenamefont{Patchkovskii,
  Zhao, Brabec, and Villeneuve}}]{patchkovskii07}
\bibinfo{author}{\bibfnamefont{S.}~\bibnamefont{Patchkovskii}},
  \bibinfo{author}{\bibfnamefont{Z.}~\bibnamefont{Zhao}},
  \bibinfo{author}{\bibfnamefont{T.}~\bibnamefont{Brabec}}, \bibnamefont{and}
  \bibinfo{author}{\bibfnamefont{D.~M.} \bibnamefont{Villeneuve}},
  \bibinfo{journal}{J. Chem. Phys.} \textbf{\bibinfo{volume}{126}},
  \bibinfo{pages}{114306} (\bibinfo{year}{2007}).

\bibitem[{\citenamefont{Altucci et~al.}(2006)\citenamefont{Altucci, Velotta,
  Heesel, Springate, Marangos, Vozzi, Benedetti, Calegari, Sansone, Stagira
  et~al.}}]{altucci1}
\bibinfo{author}{\bibfnamefont{C.}~\bibnamefont{Altucci}},
  \bibinfo{author}{\bibfnamefont{R.}~\bibnamefont{Velotta}},
  \bibinfo{author}{\bibfnamefont{E.}~\bibnamefont{Heesel}},
  \bibinfo{author}{\bibfnamefont{E.}~\bibnamefont{Springate}},
  \bibinfo{author}{\bibfnamefont{J.~P.} \bibnamefont{Marangos}},
  \bibinfo{author}{\bibfnamefont{C.}~\bibnamefont{Vozzi}},
  \bibinfo{author}{\bibfnamefont{E.}~\bibnamefont{Benedetti}},
  \bibinfo{author}{\bibfnamefont{F.}~\bibnamefont{Calegari}},
  \bibinfo{author}{\bibfnamefont{G.}~\bibnamefont{Sansone}},
  \bibinfo{author}{\bibfnamefont{S.}~\bibnamefont{Stagira}},
  \bibnamefont{et~al.}, \bibinfo{journal}{Phys. Rev. A}
  \textbf{\bibinfo{volume}{73}}, \bibinfo{pages}{043411}
  (\bibinfo{year}{2006}).

\bibitem[{\citenamefont{Lan et~al.}(2006)\citenamefont{Lan, Lu, Cao, Wang, and
  Yang}}]{lu2006}
\bibinfo{author}{\bibfnamefont{P.}~\bibnamefont{Lan}},
  \bibinfo{author}{\bibfnamefont{P.}~\bibnamefont{Lu}},
  \bibinfo{author}{\bibfnamefont{W.}~\bibnamefont{Cao}},
  \bibinfo{author}{\bibfnamefont{X.}~\bibnamefont{Wang}}, \bibnamefont{and}
  \bibinfo{author}{\bibfnamefont{G.}~\bibnamefont{Yang}},
  \bibinfo{journal}{Phys. Rev. A} \textbf{\bibinfo{volume}{74}},
  \bibinfo{pages}{063411} (\bibinfo{year}{2006}).

\bibitem[{\citenamefont{Madsen and Madsen}(2006)}]{madsen2}
\bibinfo{author}{\bibfnamefont{C.~B.} \bibnamefont{Madsen}} \bibnamefont{and}
  \bibinfo{author}{\bibfnamefont{L.~B.} \bibnamefont{Madsen}},
  \bibinfo{journal}{Phys. Rev. A} \textbf{\bibinfo{volume}{74}},
  \bibinfo{pages}{023403} (\bibinfo{year}{2006}).

\bibitem[{\citenamefont{Sundaram and Milonni}(1990)}]{MilonniD}
\bibinfo{author}{\bibfnamefont{B.}~\bibnamefont{Sundaram}} \bibnamefont{and}
  \bibinfo{author}{\bibfnamefont{P.~W.} \bibnamefont{Milonni}},
  \bibinfo{journal}{Phys. Rev. A} \textbf{\bibinfo{volume}{41}},
  \bibinfo{pages}{6571} (\bibinfo{year}{1990}).

\bibitem[{\citenamefont{Burnett et~al.}(1992)\citenamefont{Burnett, Reed,
  Cooper, and Knight}}]{BurnettHHG}
\bibinfo{author}{\bibfnamefont{K.}~\bibnamefont{Burnett}},
  \bibinfo{author}{\bibfnamefont{V.~C.} \bibnamefont{Reed}},
  \bibinfo{author}{\bibfnamefont{J.}~\bibnamefont{Cooper}}, \bibnamefont{and}
  \bibinfo{author}{\bibfnamefont{P.~L.} \bibnamefont{Knight}},
  \bibinfo{journal}{Phys. Rev. A} \textbf{\bibinfo{volume}{45}},
  \bibinfo{pages}{3347} (\bibinfo{year}{1992}).

\bibitem[{\citenamefont{Madsen et~al.}(2007)\citenamefont{Madsen, Mouritzen,
  Kjeldsen, and Madsen}}]{madsen1}
\bibinfo{author}{\bibfnamefont{C.~B.} \bibnamefont{Madsen}},
  \bibinfo{author}{\bibfnamefont{A.~S.} \bibnamefont{Mouritzen}},
  \bibinfo{author}{\bibfnamefont{T.~K.} \bibnamefont{Kjeldsen}},
  \bibnamefont{and} \bibinfo{author}{\bibfnamefont{L.~B.}
  \bibnamefont{Madsen}}, \bibinfo{journal}{arXiv:physics/0703234v1}
  (\bibinfo{year}{2007}).

\bibitem[{\citenamefont{Zare}(1988)}]{zare1}
\bibinfo{author}{\bibfnamefont{R.~N.} \bibnamefont{Zare}},
  \emph{\bibinfo{title}{Angular momentum}} (\bibinfo{publisher}{John Wiley \&
  Sons, Inc., New York}, \bibinfo{year}{1988}).

\bibitem[{\citenamefont{Stapelfeldt and Seideman}(2003)}]{stapelfeldt2003}
\bibinfo{author}{\bibfnamefont{H.}~\bibnamefont{Stapelfeldt}} \bibnamefont{and}
  \bibinfo{author}{\bibfnamefont{T.}~\bibnamefont{Seideman}},
  \bibinfo{journal}{Rev. Mod. Phys.} \textbf{\bibinfo{volume}{75}},
  \bibinfo{pages}{543} (\bibinfo{year}{2003}).

\bibitem[{\citenamefont{Rouz\'{e}e et~al.}(2006)\citenamefont{Rouz\'{e}e,
  Gu\'{e}rin, Boudon, Lavorel, and Faucher}}]{rouzee2006}
\bibinfo{author}{\bibfnamefont{A.}~\bibnamefont{Rouz\'{e}e}},
  \bibinfo{author}{\bibfnamefont{S.}~\bibnamefont{Gu\'{e}rin}},
  \bibinfo{author}{\bibfnamefont{V.}~\bibnamefont{Boudon}},
  \bibinfo{author}{\bibfnamefont{B.}~\bibnamefont{Lavorel}}, \bibnamefont{and}
  \bibinfo{author}{\bibfnamefont{O.}~\bibnamefont{Faucher}},
  \bibinfo{journal}{Phys. Rev. A} \textbf{\bibinfo{volume}{73}},
  \bibinfo{pages}{033418} (\bibinfo{year}{2006}).

\bibitem[{\citenamefont{Underwood et~al.}(2005)\citenamefont{Underwood,
  Sussman, and Stolow}}]{underwood2005}
\bibinfo{author}{\bibfnamefont{J.~G.} \bibnamefont{Underwood}},
  \bibinfo{author}{\bibfnamefont{B.~J.} \bibnamefont{Sussman}},
  \bibnamefont{and} \bibinfo{author}{\bibfnamefont{A.}~\bibnamefont{Stolow}},
  \bibinfo{journal}{Phys. Rev. Lett.} \textbf{\bibinfo{volume}{94}},
  \bibinfo{pages}{143002} (\bibinfo{year}{2005}).

\bibitem[{\citenamefont{Kuchiev and Ostrovsky}(1999)}]{kuchievandostrovsky}
\bibinfo{author}{\bibfnamefont{M.~Y.} \bibnamefont{Kuchiev}} \bibnamefont{and}
  \bibinfo{author}{\bibfnamefont{V.~N.} \bibnamefont{Ostrovsky}},
  \bibinfo{journal}{Phys. Rev. A} \textbf{\bibinfo{volume}{60}},
  \bibinfo{pages}{3111} (\bibinfo{year}{1999}).

\bibitem[{\citenamefont{Kjeldsen et~al.}(2005)\citenamefont{Kjeldsen, Bisgaard,
  Madsen, and Stapelfeldt}}]{kjeldsen1}
\bibinfo{author}{\bibfnamefont{T.~K.} \bibnamefont{Kjeldsen}},
  \bibinfo{author}{\bibfnamefont{C.~Z.} \bibnamefont{Bisgaard}},
  \bibinfo{author}{\bibfnamefont{L.~B.} \bibnamefont{Madsen}},
  \bibnamefont{and}
  \bibinfo{author}{\bibfnamefont{H.}~\bibnamefont{Stapelfeldt}},
  \bibinfo{journal}{Phys. Rev. A} \textbf{\bibinfo{volume}{71}},
  \bibinfo{pages}{013418} (\bibinfo{year}{2005}).

\bibitem[{\citenamefont{Linstrom and Mallard}(June 2005)}]{NIST}
\bibinfo{author}{\bibfnamefont{P.~J.} \bibnamefont{Linstrom}} \bibnamefont{and}
  \bibinfo{author}{\bibfnamefont{W.~G.} \bibnamefont{Mallard}},
  \emph{\bibinfo{title}{NIST Chemistry WebBook, NIST Standard Reference
  Database Number 69}} (\bibinfo{publisher}{National Institute of Standards and
  Technology, Gaithersburg MD, 20899 (http://webbook.nist.gov)},
  \bibinfo{year}{June 2005}).

\bibitem[{\citenamefont{Schmidt et~al.}(1993)\citenamefont{Schmidt, Baldridge,
  Boatz, Elbert, Gordon, Jensen, Koseki, Matsunaga, Nguyen, Su
  et~al.}}]{GAMESS}
\bibinfo{author}{\bibfnamefont{M.~W.} \bibnamefont{Schmidt}},
  \bibinfo{author}{\bibfnamefont{K.~K.} \bibnamefont{Baldridge}},
  \bibinfo{author}{\bibfnamefont{J.~A.} \bibnamefont{Boatz}},
  \bibinfo{author}{\bibfnamefont{S.~T.} \bibnamefont{Elbert}},
  \bibinfo{author}{\bibfnamefont{M.~S.} \bibnamefont{Gordon}},
  \bibinfo{author}{\bibfnamefont{J.~H.} \bibnamefont{Jensen}},
  \bibinfo{author}{\bibfnamefont{S.}~\bibnamefont{Koseki}},
  \bibinfo{author}{\bibfnamefont{N.}~\bibnamefont{Matsunaga}},
  \bibinfo{author}{\bibfnamefont{K.~A.} \bibnamefont{Nguyen}},
  \bibinfo{author}{\bibfnamefont{S.}~\bibnamefont{Su}}, \bibnamefont{et~al.},
  \bibinfo{journal}{Journal of Computational Chemistry}
  \textbf{\bibinfo{volume}{14}}, \bibinfo{pages}{1347} (\bibinfo{year}{1993}).

\bibitem[{\citenamefont{Priori et~al.}(2000)\citenamefont{Priori, Cerullo,
  Nisoli, Stagira, Silvestri, Villoresi, Poletto, Ceccherini, Altucci, Bruzzese
  et~al.}}]{priori1}
\bibinfo{author}{\bibfnamefont{E.}~\bibnamefont{Priori}},
  \bibinfo{author}{\bibfnamefont{G.}~\bibnamefont{Cerullo}},
  \bibinfo{author}{\bibfnamefont{M.}~\bibnamefont{Nisoli}},
  \bibinfo{author}{\bibfnamefont{S.}~\bibnamefont{Stagira}},
  \bibinfo{author}{\bibfnamefont{S.~D.} \bibnamefont{Silvestri}},
  \bibinfo{author}{\bibfnamefont{P.}~\bibnamefont{Villoresi}},
  \bibinfo{author}{\bibfnamefont{L.}~\bibnamefont{Poletto}},
  \bibinfo{author}{\bibfnamefont{P.}~\bibnamefont{Ceccherini}},
  \bibinfo{author}{\bibfnamefont{C.}~\bibnamefont{Altucci}},
  \bibinfo{author}{\bibfnamefont{R.}~\bibnamefont{Bruzzese}},
  \bibnamefont{et~al.}, \bibinfo{journal}{Phys. Rev. A}
  \textbf{\bibinfo{volume}{61}}, \bibinfo{pages}{063801}
  (\bibinfo{year}{2000}).

\bibitem[{\citenamefont{Tosa et~al.}(2003)\citenamefont{Tosa, Takahashi,
  Nabekawa, and Midorikawa}}]{tosi1}
\bibinfo{author}{\bibfnamefont{V.}~\bibnamefont{Tosa}},
  \bibinfo{author}{\bibfnamefont{E.}~\bibnamefont{Takahashi}},
  \bibinfo{author}{\bibfnamefont{Y.}~\bibnamefont{Nabekawa}}, \bibnamefont{and}
  \bibinfo{author}{\bibfnamefont{K.}~\bibnamefont{Midorikawa}},
  \bibinfo{journal}{Phys. Rev. A} \textbf{\bibinfo{volume}{67}},
  \bibinfo{pages}{063817} (\bibinfo{year}{2003}).

\bibitem[{\citenamefont{Levesque et~al.}(2007)\citenamefont{Levesque, Zeidler,
  Marangos, Corkum, and Villeneuve}}]{levesque1}
\bibinfo{author}{\bibfnamefont{J.}~\bibnamefont{Levesque}},
  \bibinfo{author}{\bibfnamefont{D.}~\bibnamefont{Zeidler}},
  \bibinfo{author}{\bibfnamefont{J.~P.} \bibnamefont{Marangos}},
  \bibinfo{author}{\bibfnamefont{P.~B.} \bibnamefont{Corkum}},
  \bibnamefont{and} \bibinfo{author}{\bibfnamefont{D.~M.}
  \bibnamefont{Villeneuve}}, \bibinfo{journal}{Phys. Rev. Lett.}
  \textbf{\bibinfo{volume}{98}}, \bibinfo{pages}{183903}
  (\bibinfo{year}{2007}).

\bibitem[{\citenamefont{Chiril\u{a} and Lein}(2006)}]{lein2006}
\bibinfo{author}{\bibfnamefont{C.~C.} \bibnamefont{Chiril\u{a}}}
  \bibnamefont{and} \bibinfo{author}{\bibfnamefont{M.}~\bibnamefont{Lein}},
  \bibinfo{journal}{Phys. Rev. A} \textbf{\bibinfo{volume}{73}},
  \bibinfo{pages}{023410} (\bibinfo{year}{2006}).

\bibitem[{\citenamefont{Chen and Chen}(2007)}]{chen2007}
\bibinfo{author}{\bibfnamefont{J.}~\bibnamefont{Chen}} \bibnamefont{and}
  \bibinfo{author}{\bibfnamefont{S.~G.} \bibnamefont{Chen}},
  \bibinfo{journal}{Phys. Rev. A} \textbf{\bibinfo{volume}{75}},
  \bibinfo{pages}{041402(R)} (\bibinfo{year}{2007}).

\end{thebibliography}

\end{document}